\documentclass[journal]{IEEEtran}

\usepackage{hyperref}
\usepackage{amsmath,amssymb,amsthm}
\usepackage{graphicx}
\usepackage{epstopdf, xspace}%,cite}

\usepackage{amsthm}
\usepackage{amssymb,amsmath}
\usepackage{amsmath}
\usepackage{graphicx}
\usepackage{hyperref}% http://ctan.org/pkg/hyperref
\usepackage{cleveref}
\usepackage{lipsum}
\usepackage{epsfig}
\usepackage{url}
\usepackage{caption}
\usepackage{float}
\usepackage{tikz}

\usepackage{enumitem}
\usepackage{latexsym,color}
\usepackage{xspace}

\newcommand{\ie}{\textit{i.e.,}\xspace}
\newcommand{\eg}{\textit{e.g.,}\xspace}

\newtheorem{theorem}{Theorem}

\newtheorem{lemma}{Lemma}
\newtheorem{definition}{Definition}

\usepackage{subfigure}
\usepackage{caption}

\begin{document}

\title{On the Approximability of Related Machine Scheduling under Arbitrary Precedence}

\author{
	Vaneet Aggarwal, Tian Lan, Suresh Subramaniam, and Maotong Xu\thanks{V. Aggarwal is with the  School of Industrial Engineering and the School of Electrical and Computer Engineering, Purdue University, West Lafayette IN 47907, email: vaneet@purdue.edu. T. Lan, S.  Subramaniam, and M. Xu are with the Department of ECE,
			The George Washington University, DC 20052, USA, email: \{tlan, suresh, htfy8927\}@gwu.edu. 
	This work was supported in part by ONR Grant N00014-20-1-2146, U.S. National Science Foundation under Grant  CNS-1618335, and CISCO research gift 1155109.	
	} } 
	%\\
	%{$^{1}$Department of ECE,
	%	The George Washington University, DC 20052, USA} \\
%{$^{2}$School of IE 
%		Purdue University, West Lafayette, IN 47907, USA} \\
%	 vaneet@purdue.edu, tlan@gwu.edu, suresh@gwu.edu, htfy8927@gwu.edu
%}

\maketitle

\begin{abstract}
Distributed computing systems often need to consider the scheduling problem involving a collection of highly dependent data-processing tasks that must work in concert to achieve mission-critical objectives. This paper considers the unrelated machine scheduling problem for minimizing weighted sum completion time under arbitrary precedence constraints and on heterogeneous machines with different processing speeds. The problem is known to be strongly NP-hard even in the single machine setting. By making use of Queyranne's constraint set and constructing a novel Linear Programming relaxation for the scheduling problem under arbitrary precedence constraints, our results in this paper advance the state of the art. We develop a $2(1+(m-1)/D)$-approximation algorithm (and $2(1+(m-1)/D)+1$-approximation) for the scheduling problem with zero release time (and arbitrary release time), where $m$ is the number of servers and $D$ is the task-skewness product. The algorithm can be efficiently computed in polynomial time using the Ellipsoid method and achieves nearly optimal performance in practice as $D>O(m)$ when the number of tasks per job to schedule is sufficiently larger than the number of machines available. Our implementation and evaluation using a heterogeneous testbed and real-world benchmarks confirms significant improvement in weighted sum completion time for dependent computing tasks.
%MapReduce is the most popular big-data computation framework, motivating many research topics. A MapReduce job consists of two successive phases, \ie map phase and reduce phase. Each phase can be divided into multiple tasks. A reduce task can only start when  all the map tasks finish processing. A job is successfully completed when all its map and reduce tasks are complete. The task of optimally scheduling the different tasks on different servers to minimize the weighted completion time is an open problem, and is the focus of this paper. In this paper, we give an approximation ratio with a competitive ratio $2(1+(m-1)/D)+1$, where $m$ is the number of servers and $D\ge 1$ is the task-skewness product. We implement the proposed algorithm on Hadoop framework, and compare with three baseline schedulers. Results show that our DMRS algorithm can outperform baseline schedulers by up to $82\%$.
\end{abstract}

%\thispagestyle{empty}

%\input{abstract}
%\the\textwidth; \the\textheight
\begin{IEEEkeywords}Related Machine Scheduling, Precedence Constraints, Directed Acyclic Graph, Approximation Algorithm.
\end{IEEEkeywords}

%\vspace{-.1in}
\section{Introduction}\label{sec:introduction}

%MPI~\cite{mpi}

%Big Data has emerged in the past few years as a new paradigm presenting abundant oppurtunities and challenges, including efficient processing and computation of data. This has led to the development of parallel computing frameworks, such as MapReduce~\cite{Dean:2008}, designed to process massive amounts of data. MapReduce has two fundamental processes, {\em map} and {\em reduce}. Input data are first split into smaller segments that are processed by parallel map tasks on different machines. The intermediate output, consisting of key-value pairs, are then processed by reduce tasks to obtain the final result. Due to the increasing level of heterogeneity in both application requirements and computing infrastructures, scheduling algorithms for MapReduce framework have been widely studied with the goal of reducing job completion times \cite{chen2010samr,tang2013mapreduce,thomas2014survey,HuangBWLCT15,Wang:2016}.

Next-generation computing systems such as distributed learning are becoming increasingly sophisticated and heterogeneous, often involving a collection of highly dependent data-processing tasks that must work in concert to achieve mission-critical objectives, going beyond traditional considerations like throughput or congestion. For instance, %in distributed deep learning with synchronous training \cite{DBLP:conf/icml/TandonLDK17}, gradients are computed on decentralized workers before being transferred to an aggregator for model update, while processing orders are also required between any consecutive gradient steps. 
data processing frameworks like MapReduce execute tasks in multiple sequential stages. Visual Question Answering (VQA) applications \cite{Cadene_2019_CVPR} often perform multiple steps of reasoning and processing, each time refining the representations with contextual and question information. In general, such precedence constraints that exist in distributed computing can be formulated as a partial order among all tasks belonging to the same job, i.e., $i\preceq j$ if task $i$ must be completed before task $j$ starts. { The problem of task scheduling with precedence constraints arises in multi-cloud environments \cite{panda2015efficient,panda2018normalization,masdari2020efficient}, where the precedence constraint is important to consider for scheduling on related servers, and is the subject of this paper. }

\begin{figure}[!t]
	\centering
	\includegraphics[width=0.4\textwidth]{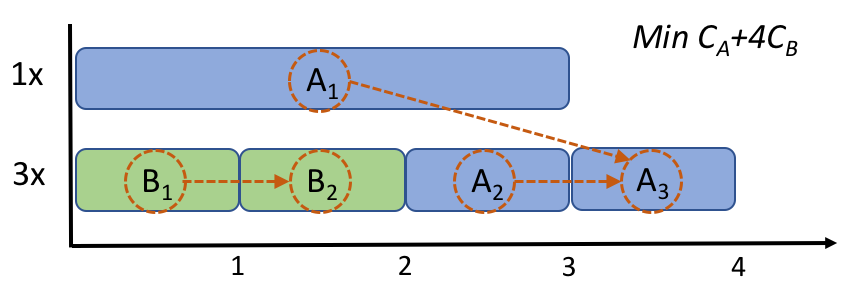}
	\caption{\small Related machine scheduling under precedence constraints is NP-hard even in special cases. Algorithms that either assume identical machine speed or are oblivious of such dependence would result in substantially higher weighted completion time than the optimal solution shown here.}
	\label{fig:example}
	\vspace{-.15in}
\end{figure}

In this paper, we consider the scheduling problem of minimizing weighted completion time of multiple learning jobs under precedence constraints (that are modeled as an arbitrary Directed Acyclic Graph (DAG), or dependence graph) and on heterogeneous machines (with different processing speeds). Each vertex $i$ in the dependence graph denotes a job, while each arc $(i,j)$ represents a precedence constraint $i\preceq j$ between jobs $i$ and  $j$, i.e., all tasks of job $i$ must be completed before any task of job $j$ starts. A job is completed if all its constituent tasks are finished. An example of scheduling problem is illustrated in Figure~\ref{fig:example}, with two jobs $\{A_1,A_2,A_3\}$ and $\{B_1,B_2\}$ under precedence constraints between their constituent tasks. { The weighted completion time is $w_A C_A + w_B C_B$, where $C_A$ and $C_B$ are the completion times of the jobs $A$ and $B$, respectively. In Fig. \ref{fig:example}, we assume $w_A=4$ and $w_B=1$. }  It is easy to see that an optimal solution minimizing weighted completion time  must take precedence constraints into account, while algorithms that either assume identical machine speed or are oblivious of such dependence would result in substantially higher weighted completion time. { The machines are assumed to have different speeds, e.g., speed of second machine is three times that of the first machine in Fig. \ref{fig:example}. }  %Note that we do not consider time-slot based scheduling algorithms allowing fractional job executions as in \cite{Chudak:1997,Li:2017}. 
 This scheduling problem with different machine speeds and under precedence constraints has been studied dating back to the 1950s \cite{smith1956various}, but still remains an open problem despite recent progress on approximation algorithms in a few special cases \cite{ChangKKLLM11,skutella2001convex,schuurman1999polynomial,im2016better,sviridenko2013approximating,murray2016scheduling,bansal2016lift}.  When  there is a single job, thus giving no precedence constraints, the problem of scheduling jobs on machines with different processing speed has been studied as the {\em Related Machine Scheduling} problem, which is known to be strongly NP-hard even in the single machine setting, and APX-hard even when all jobs
are available to schedule at time 0 (referred to as zero release time) \cite{im2016better}. 
%Different approximation algorithms have been proposed for the problem in \cite{ChangKKLLM11,skutella2001convex,schuurman1999polynomial,im2016better,sviridenko2013approximating,murray2016scheduling,bansal2016lift}. 
%This problem is listed in \cite{schuurman1999polynomial} as one of the top ten open problems in the field of approximatescheduling algorithms, and 
The best known result is a 1.5-approximation algorithm for zero release times \cite{bansal2016lift}, and 1.8786-approximation algorithm for arbitrary arrival times \cite{im2016better}. Later, for the special case of identical machines and multiple jobs, and when the dependence graph reduces to a complete bipartite graph, 3-approximation and 7-approximation algorithms are proposed in \cite{ChenKL12,YuanWL14} for zero and general release times, respectively. %However, the best known scheduling algorithms with arbitrary machine speed - and yet bipartite dependence graphs - are far from optimal, with 54-approximation \cite{fotakis2015scheduling} and 37.86-approximation \cite{fotakis2016scheduling}. 

{Our results in this paper advance the state of the art on related machine scheduling under (i) arbitrary precedence constraints (i.e., any dependence graph) and (ii) heterogeneous machine speeds, when each job consists of multiple parallel tasks.} In particular, we consider the problem of minimizing weighted sum completion time $\sum_s w_sC_s$, where $C_s$ denotes the completion time of job $s$, which is determined by the completion of all its constituent tasks, and $w_s$ is a non-negative weight for job $s$. We develop a $2(1+(m-1)/D)$-approximation algorithm for the scheduling problem with zero release time, where $m$ is the number of machines and $D$ is a metric quantifying the task-skewness product, which is defined as the minimum (over all jobs in a set) ratio of the sum of task sizes (in a job) to the largest task size (in that job). Since the number of tasks to schedule is normally much larger than the number of machines available, we have $D>O(m)$, which implies that our proposed algorithm achieves an approximation ratio of $2+\epsilon$ in practice. 
%when all jobs are released at time 0, where $m$ is the number of machines and $D$ is a metric quantifying the task-skewness product of map and reduce tasks, defined as the sum processing size of all map and reduce tasks divided by that of the largest map and reduce tasks Thus, we have $D\ge 1$, and the value of $D$ increases with increased number of tasks, and is also higher for a  left-skewed task-size distribution (\ie  a larger percentage of large-size jobs). 
Further, we show that the competitive ratio becomes $2(1+(m-1)/D)+1$ for general job release times, or $3+\epsilon$ when $D>O(m)$. The key idea of our approach is to make use of the Queyranne's constraint set \cite{queyranne1993structure} and construct a novel Linear Programming (LP) relaxation for the scheduling problem under arbitrary precedence constraints. 
%schedule map and reduce tasks by solving an approximated linear program, in which the dependence between map and reduce tasks is cast into (precedence) constraints with respect to map/reduce completion times. 
Then, we show that the proposed LP can be efficiently computed in polynomial time using the Ellipsoid method \cite{grotschel1981ellipsoid}. It yields a scheduling algorithm with provable competitive ratio with respect to the optimal scheduling solution. Even when restricted to complete bipartite dependence graphs, our results significantly improve prior work, namely 37.86-approximation algorithm proposed in \cite{fotakis2016scheduling} ({ though their result is for a more general setting of unrelated machines}, while only for zero release times), and achieves nearly optimal performance when the number of tasks to schedule is sufficiently larger than the number of machines available.  {We also note that 54-approximation algorithm proposed in \cite{fotakis2015scheduling} is for the case of disjoint processors in bi-partite graphs, where the Map and Reduce jobs do not happen on the same servers, and thus the setup is not directly comparable. In addition, we do not consider preemption of jobs like in \cite{moseley2011scheduling}. The setting of related machine scheduling is studied in \cite{Chudak:1997,Li:2017}, where approximation guarantee in \cite{Li:2017} is $O(\log m/\log \log m)$. For $D>O(m\log \log m/\log m)$, we outperform the state of the art results in \cite{Li:2017}. } The results are compared in Table~I, {which includes a summary of different results for identical and unrelated machines.} %We note that for bipartite $(n_1,n_2)$ graph can be represented with two jobs and for equal task sizes in each job, we have $D=\min(n_1,n_2)$. Thus, for large $n_1$ and $n_2$, $D$ can be made large enough. 

We implement the proposed scheduling algorithm and evaluate its performance in Hadoop, whose map and reduce tasks satisfy a complete bipartite dependence graph.  Modifications to the Application Master and Resource Manager are made to ensure that task/job execution follow the desired order, as given by our optimization solution. Our extensive experiments, using a combination of WordCount, Sort, and TeraSort benchmarks on a heterogeneous cluster and on real-world datasets (resulting in high level of execution time uncertainty), validate that our proposed scheduler outperforms baseline strategies including FIFO in default Hadoop, Identical-machine \cite{ChenKL12,YuanWL14}, and Map-only \cite{im2016better}, in terms of sum weighted completion time. Especially, the scheduler can achieve the smallest total weighted completion time for scheduling benchmarks with heavy workloads in reduce phase, \eg TeraSort. We also perform simulations to evaluate our scheduler under dependence graphs with multiple waves of execution. We note that for equal-sized map and equal-sized reduce jobs, $D$ will be the number of map-reduce jobs and thus the algorithm would work in $D>O(m)$ regime.

%A key feature of the implementation is its ability to adapt to system execution dynamics and uncertainty. In particular, we implement a task scheduler that not only computes an optimal schedule of map and reduce tasks according to the proposed solution, but also has the ability to re-optimize the schedule on the fly based on available task progress and renewed completion time estimates. We also modify the Application Master and Resource Manager in Hadoop to ensure the task/job execution in the desired order, as well as to handle potential desynchronization and disconnection issues. Our extensive experiments, using a combination of WordCount, Sort, and TeraSort benchmarks on a heterogeneous cluster, validate that our proposed DMRS outperforms FIFO, Identical-machine, and Map-only, in terms of total weighted completion time. Especially, DMRS can achieve the smallest total weighted completion time for scheduling benchmarks with heavy workloads in reduce phase, \eg TeraSort.

The main contributions of the paper are as follows. 
\begin{itemize}
	\item We consider the related machine scheduling problem for minimizing weighted sum completion time, under arbitrary precedence constraints and on heterogeneous machines. 
	\item The proposed scheduling algorithm is based on the solution of an approximated linear program, which recasts the precedence constraints and is shown to be solvable in polynomial time. 
	\item We analyze the proposed scheduling algorithm and quantify its approximation ratio with both zero and arbitrary release times, which significantly improves prior art, especially when the number of tasks per job is large.
	\item Our implementation and evaluation using Hadoop shows that the scheduler outperforms other baselines by up to $82\%$ in terms of total weighted completion time. 
	%\item We evaluate proposed DMRS through large-scale simulations. Results show that DMRS outperforms other scheduling algorithms, e.g., High Unit Weight First, Tetris~\cite{grandl2015multi}, and FIFO.
\end{itemize}
The rest of the paper is organized as follows. We present the system model and formulate the problem in Section~\ref{sec:system_model}. The approximation algorithm and its analysis are provided in Section~\ref{sec:approximation_algorithm}. The implementation details of the proposed algorithm are provided in Section \ref{sec:implementation}, and evaluation results are presented in Section~\ref{sec:evaluation}. %Section~\ref{sec:evaluation} provides the experimental evaluation and simulation results.

%\input{1_background.tex}
%\vspace{-.1in}

\begin{table*}[htp!]
\centering
\resizebox{\textwidth}{!}{%
\begin{tabular}{  | l l l c | c|}
\hline 
\multicolumn{4}{|c|}{ {\bf Scheduling problems} } & {\bf Performance bounds}  \\
\hline
Arbitrary machines;  & Single-vertex graphs; & Arbitrary release time &  \cite{im2016better} &  $1.8786$  \\
\hline
Identical machines;  & Bipartite graphs; & Zero release time & \cite{ChenKL12} &  $3$  \\
\hline
Identical machines;  & Bipartite graphs; & Arbitrary release time & \cite{YuanWL14} &  $7$  \\
\hline
Identical machines;  & Arbitrary graphs; & Zero release time & \cite{fotakis2016scheduling} &  4%\footnote{The result holds for constants $\alpha>1$ and $\epsilon>0$, and reduces to 37.87 for $k=2$, i.e., bipartite graphs. }  
\\
\hline
Identical machines;  & Arbitrary graphs; & Zero release time & \cite{Li:2017} &  $2+2\log 2 +\epsilon$ %\footnote{The result holds for constants $\alpha>1$ and $\epsilon>0$, and reduces to 37.87 for $k=2$, i.e., bipartite graphs. }  
\\
\hline
%Arbitrary machines;  & Chains; & Arbitrary release time & \cite{Woeginger:2000} &  $2$  \\
%\hline
Unrelated machines, disjoint processors;  & Bipartite graphs; & Zero release time & \cite{fotakis2015scheduling} &  $54$  \\
\hline
Unrelated machines;  & $k$-partite graphs\footnotemark[1]; & Zero release time & \cite{fotakis2016scheduling} &  $(k+1+\frac{k}{\epsilon})\frac{(2\alpha^2-1)(1+\epsilon)}{\alpha-1}+\alpha+\frac{\alpha}{\epsilon}$
\\
%\hline
%Arbitrary machines;  & Chains; & Arbitrary release time & \cite{Woeginger:2000} &  $2$  \\
\hline
%Unrelated machines, disjoint processors;  & Bipartite graphs; & Zero release time & \cite{fotakis2015scheduling} &  $54$  \\
%\hline
Related machines;  & Arbitrary graphs; & Zero release time & \cite{Chudak:1997} &  $O(\log m)$\\
%\footnote{The result holds for constants $\alpha>1$ and $\epsilon>0$, and reduces to 37.87 for $k=2$, i.e., bipartite graphs. }  
%\\
\hline
Related machines;  & Arbitrary graphs; & Zero release time & \cite{Li:2017} &  $O(\log m/\log \log m)$\\
%\footnote{The result holds for constants $\alpha>1$ and $\epsilon>0$, and reduces to 37.87 for $k=2$, i.e., bipartite graphs. }  
%\\
\hline
Related machines; & Arbitrary graphs; & Zero release time & [This paper] &  $2(1+\frac{m-1}{D})$, or $(2+\epsilon)$ as {\tiny $D>O(m)$}  \\
\hline
Related machines; & Arbitrary graphs; & Arbitrary release time & [This paper]   & $1+2(1+\frac{m-1}{D})$, or $(3+\epsilon)$ as {\tiny $D>O(m)$} \\
\hline
\end{tabular} } \label{tbl:ex}%\vspace{.1in}
\caption{Comparing performance bounds of related machine scheduling under different conditions and precedence constraints (i.e., dependence graphs), for minimizing weighted completion times $\sum_iw_iC_i$. \\ \vspace{-0.in} {\small $^1$ The bound holds for $\alpha>1$ and $\epsilon>0$, and reduces to 37.87 for $k=2$, i.e., bipartite graphs.} }
\vspace{-.15in}
\end{table*}
%\footnotetext[1]{The bound holds for $\alpha>1$ and $\epsilon>0$, and reduces to 37.87 for $k=2$, i.e., bipartite graphs.}

\section{System Model and Problem Formulation} \label{sec:system_model}

We consider scheduling $N$ jobs, where the set of jobs is denoted by $\mathcal{J}$. Each job $j\in \mathcal{J}$ consists of $t_j$ tasks, where the set of tasks of job $j$ is denoted by ${\mathcal T}_j$. Task $t\in {\mathcal T}_j$ of job $j$ has processing data size $p_{j,t}$. Without loss of generality, we assume that the tasks of a job are ordered in a non-increasing order of the task sizes, or $$p_{j,1}\geq p_{j,2}\geq \cdots \geq p_{j,t_j}.$$

The objective of the problem considered in this paper is to schedule different tasks on $m$ heterogeneous machines. Further, the scheduling is assumed to be non-preemptive, i.e., once a task is started on a machine, it cannot be stopped until it is complete. We assume that the speed of machine $i\in\{1,\cdots,m\}$ is $v_i$. The  time taken for processing task $t\in T_j$ on server $i$ is $p_{j,t}/v_i$. Without loss of generality, we assume that machines are ordered in non-increasing order of their speeds, or $$v_1\geq v_2\geq \cdots \geq v_m.$$

We assume that the different tasks in a job have no constraints, and thus can be scheduled in parallel. The set of jobs  have precedence constraints which can be represented by a directed acyclic graph $G=(V,E)$, where every node represents a job. Every directed edge $(j_1,j_2)\in E$ is a constraint that job $j_2$ cannot start until the completion of  job $j_1$. In other words, no task of $j_2$ can start before all tasks in $j_1$ are completed. The graph is assumed to be acyclic since there is no possible ordering of jobs that satisfies the precedence constraints if there is a cycle. Every job $j$ has release time $r_j$, and has weight $w_j$. Without loss of generality, we assume that if $(j_1,j_2)\in E$, then $r_{j_2}\geq r_{j_1}$. This is because the start time of every task in $j_2$ is after the completion time of all tasks in $j_1$ which is at least $r_{j_1}$. 
The aim is to minimize the weighted completion time of the jobs. Let $C_j^{OPT}$ represent the  completion time of job $j$ based on the scheduling of different tasks, such that  $\sum_j w_j C_j$ is minimized, where $C_j$ is the completion time of job $j$.

% which can be scheduled in parallel.  

%Suppose we have a job set $\mathcal{J}$ containing $N$ jobs, each job $j$ contains a set $T_j$ of $t_j$ tasks to be assigned on $m$ machines, every task $t\in T_j$ of job $j$ has size $p_{j,t}$ (expected size $\mathbb{E}[p_{j,t}]$), and every machine $i\in\{1,\cdots,m\}$ has speed $v_i$. Without loss of generality, we assume that $v_i$ are sorted in nonincreasing order, namely $$v_1\geq v_2\geq \cdots \geq v_m.$$ Without loss of generality, we assume that $\mathbb{E}[p_{j,t}]$ are sorted in non-increasing order, namely $$\mathbb{E}[p_{j,1}]\geq \mathbb{E}[p_{j,2}]\geq \cdots \geq \mathbb{E}[p_{j,T_j}].$$  The expected time for processing task $t\in T_j$ on server $i$ is $\mathbb{E}[p_{j,t}]/v_i$. Every job $j$ has release time $r_j$, and weight $w_j$. Different tasks of the same job can be processed on different machines at the same time. All of the tasks are scheduled without preemption: once a task is started on one machine, it cannot be stopped processing until its completion.

%The set of jobs also have precedence constraints which can be denoted as a directed acyclic graph $G=(V,E)$, where every node represents a job. Every edge $(j_1,j_2)\in E$ is a constraint that job $j_2$ cannot start until the completion of of job $j_1$. 

We introduce some notations that are employed in this paper to simplify the analysis and discussions. Let $\mu$ denote the total processing rates of $m$ machines, i.e., $\mu=\sum_{l=1}^m v_l$. Further, let $q_j$  denote the maximum number of concurrent tasks in a job, i.e., $q_j = \min\{t_j,m\}$, which can be scheduled in parallel. We define $\mu_j$ to be the sum processing speed of the fastest $q_j$ machines,  or  $\mu_j=\sum_{l=1}^{q_j} v_l$. It is easy to see that $\mu_j$ is the maximum possible processing speed of job $j$,
since its tasks can only occupy $q_j$ distinct machines at any given time. We denote the total processing data size of all  tasks of job $j$ as $p_j$, i.e.,  $p_j = \sum_{t\in T_{j}} p_{j,t}$. %Thus, $\mathbb{E}[p_{j}]=\sum_{t\in {T_j}} \mathbb{E}[p_{j,t}]$ is the expected total processing data size of tasks in job $j$. 

We define the task-skewness product of a job as the ratio of  the total  size of job $j$ and the maximum  task size in job $j$, i.e., $D_j={p_{j}}/{p_{j,1}}$. This can also be seen as the number of tasks in a job times the average-to-max ratio of the task sizes of the job. Thus, if there are multiple tasks of equal size, $D$ equals the number of tasks in the job. %$D$ is larger when there are more larger-sized tasks in a job (left-skewed).
 The task-skewness product of all jobs,  denoted by $D$, is the minimum of $D_j$ for all $j\in \mathcal{J}$, i.e., $D = \min_j {p_{j}}/{p_{j,1}}$. %Task-skewness is independent of the number of machines, 

\section{Approximation Algorithm} \label{sec:approximation_algorithm}
%\vspace{-.05in}

In this section, we develop an algorithm, S-PC (Scheduling under Precedence Constraints), to solve the weighted   completion time minimization problem on heterogeneous machines  and under precedence constraints. The algorithm is based on first solving a linear optimization, referred to as the LP-Schedule problem. The solution is then used to obtain a feasible schedule for executing the tasks on the machines.

%\vspace{-.15in}
\subsection{LP-Schedule}
%\vspace{-.05in}

We formulate LP-Schedule as follows:
%\vspace{-.05in}
\begin{equation}
 \min \sum_{j=1}^N w_jC_j \label{eq1}
\end{equation}
\begin{align}
& \rm{s.t.} && \sum_{j\in S}p_jC_j\geq f(S),\quad\forall S\subset\{1,\cdots,N\};\label{eq2}\\
& &&C_{j}\geq p_j/\mu_j+r_j \qquad \forall j\in\{1,\cdots,N\};\label{eq3}\\
& && C_j\geq p_j/\mu_j+C_{j'}\qquad \forall j\in\{1,\cdots,N\}, (j',j)\in E\label{eq4}
\end{align}
where 
\begin{equation*}
f(S)=\sum_{j\in S}\frac{1}{2\mu_j}\big{(}p_j\big{)}^2+\frac{1}{2\mu}\Big{(}\sum_{j\in S}p_j\Big{)}^2.
\end{equation*}

In this formulation, $C_j$ represents the  completion time of job $j$.  We note that constraint \eqref{eq2} is based on the Queyranne's constraint set \cite{queyranne1993structure}, which has been used to give 2-approximation for concurrent open shop scheduling \cite{leung2007scheduling,mastrolilli2010minimizing} without precedence constraints. The extension to machines with different processing speeds is due to \cite{murray2016scheduling}, which formulated different versions of the polyhedral constraints based on Queyranne's constraint set.   %We note that these works assume deterministic processing times for the tasks. 
It was shown in \cite{murray2016scheduling} that this constraint is necessary for deterministic processing times.% while the same is not true for stochastic processing times. Indeed, the authors of \cite{mohring1999approximation} show that the constraint is not necessary for stochastic processing times, even for identical machines. 

%In contrast, the proposed constraint in \eqref{eq2} works on stochastic task sizes. In the following result, we will show that \eqref{eq2} gives a necessary constraint for the considered problem. 

Constraint \eqref{eq3} means that the completion time of a job is at least the sum of the release time and the processing time of the job on the fastest $q_j$ machines. Constraint \eqref{eq4} replaces the release time in \eqref{eq3} by the completion time of the jobs which have to finish prior to job $j$ based on the precedence constraints.  It is easy to see that these constraints are also necessary for the proposed problem.

Because constraints \eqref{eq2}-\eqref{eq4} are necessary for any feasible solution of the weighted   completion time optimization, any optimal solution of the LP-Schedule provides a lower bound for the weighted   completion time optimization. This lower bound may not be tight, and the optimal solution may not be feasible in the original formulation, since LP-Schedule does not take into account all sufficient constraints. Nevertheless, we show that a feasible schedule for executing different tasks on different machines can be obtained from the optimal solution of the LP-Schedule. %We note that such feasible schedule also holds for stochastic processing times. 

%all map tasks is at least the release time (i.e., the earliest time job $j$ can begin processing) plus the required processing time of any map task t on the fastest machine (i.e., the minimum required processing time of any map task).

%Constraint \eqref{eq4} describes all precedence constraints. If $(j',j)\in E$, job $j$ can only start after the completion of job $j'$. Therefore, the expected completion time of job $j$ is at least the expected completion time of job $j'$ plus the shortest expected processing time of job $j$.

%\vspace{-.3in}
\subsection{Complexity of Solving LP-Schedule}
%\vspace{-.05in}
Note that \eqref{eq2} contains an exponential number of inequalities. We are still able to solve the linear programs in polynomial time with the Ellipsoid method by using a similar separation oracle as in \cite{murray2016scheduling}.
%\vspace{-.05in}
\begin{definition}[Oracle LP-Schedule]
	Define the violation as
	$$V(S)=\frac{1}{2}\Big{[}\frac{(\sum_{j\in S} p_j)^2}{\mu}+\frac{\sum_{j\in S} (p_j)^2}{\mu_j}\Big{]}-\sum_{j\in S}p_jC_j.$$
	Let $\{C_j\}\in \mathbb{R}^N$ be a potentially feasible solution to our LP-Schedule \eqref{eq1}-\eqref{eq4}. Let $\tau$ denote the ordering when jobs are sorted in increasing order of $C_j-p_j/(2\mu_j)$. Find the most violated constraint in \eqref{eq2} by searching over $V(S)$ for $S$ of the form $\{\tau(1),\cdots,\tau(j)\}$. If maximal $V(S^*)>0$, return $S^*$ as a violated constraint for \eqref{eq2}. Otherwise, check \eqref{eq3}-\eqref{eq4} in linear time.
\end{definition}

The difficulty of testing whether a set of job completion times is a feasible solution to \eqref{eq1}-\eqref{eq4} is to find the most violated constraint in \eqref{eq2} for any subset of jobs. The above definition only provides us some set of jobs of the form $\{\tau(1),\cdots,\tau(j)\}$ (where $j\in\{1,\cdots,N\}$) instead of all subsets of jobs. The following lemma shows that it is sufficient to guarantee that any choice of jobs does not violate \eqref{eq2} once every set of jobs in the form $\{\tau(1),\cdots,\tau(j)\}$ (where $j\in\{1,\cdots,N\}$) does not violate \eqref{eq2}.

\begin{lemma}[Special case of Lemma 5, \cite{murray2016scheduling}]
	For $P(A):=\sum_{j\in A}p_j$, we have $x\in S^* \Leftrightarrow C_x-p_x/(2\mu_x)\leq P(S^*)/\mu$.
\end{lemma}
Based on this lemma, we can find $S^*$ in $O(n\log (n))$ time (sorting $C_{\tau(j)}-p_{\tau(j)}/2\mu_{\tau(j)}$ and compute each set), which implies that Oracle LP-Schedule runs in $O(n\log (n))$ time. By the equivalence of separation and optimization, we have  the following result.
\begin{theorem}
	LP-Schedule can be solved in polynomial time.
\end{theorem}

%\vspace{-.2in}
\subsection{Proposed Algorithm}

The proposed algorithm for  scheduling with precedence constraints on heterogeneous machines is denoted S-PC, and is described in Algorithm~1. This algorithm is based on list scheduling. 

We first solve the LP-relaxation problem \eqref{eq1}-\eqref{eq4} using the release times, job weights, machine speeds, directed acyclic graph for job precedence, and the  task processing times. We assume that the solution of $C_j$ from \eqref{eq1}-\eqref{eq4} is $C_j^{LP}$ for all $j\in\{1,\cdots,N\}$, and the jobs $\sigma(1),\cdots, \sigma(N)$ are sequenced in non-decreasing order of $C_{\sigma(j)}^{LP}$. Thus, the tasks of job $\sigma(i)$ are scheduled before that of $\sigma(j)$ for $i<j$. In order to schedule  different tasks in a job, we use Algorithm~2.

Note that different tasks in a job are ordered in non-increasing order of  processing data sizes. We schedule the tasks of a job in order. Each task is assigned  to a machine that produces the earliest completion time, with respect to all tasks already assigned to the machine, the task $t$'s release time, and the required processing speed of task $t$ on the machine. The detailed procedure can be seen in Algorithm~2.

Given the above procedure, we obtain a list of an ordering of all tasks on each machine. The tasks on each machine are run in the order decided. We will insert idle times on all machines as necessary, if any job $j$'s  tasks have an earlier starting time than the completion of tasks of job $i$ for $(i,j)\in E$ (if there is a precedence constraint), and if any jobs are ``scheduled'' to begin processing ahead of their release times. This procedure ensures that job precedence constraints and job release time constraints are satisfied.

\begin{figure}[!thp]
	%\subfigure[Comparing storage capacity] {
		\includegraphics[width=0.5\textwidth]{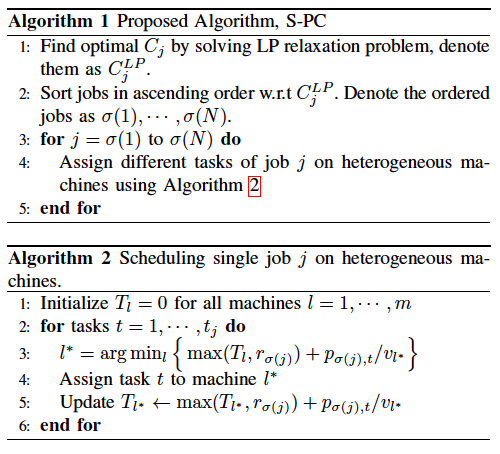}
		\vspace{-0.1in}
\end{figure}
%\input{3_approximation_algorithm.tex}
%\vspace{-.1in}

\section{Proof of S-PC Approximation Ratio }\label{sec:guarantee}

In this section, we will provide approximation guarantees for the proposed algorithm. We first note that since LP-Schedule has the original objective with necessary constraints, we have
\begin{equation}
\sum_j w_j C_j^{LP} \le \sum_j w_j C_j^{OPT}, \label{lplessopt}
\end{equation}
where $C_j^{OPT}$ is the  completion time of job $j$ with optimal scheduling. Let $\widehat{C}_j$ be the  completion time of job $j$ using Algorithm~1, and $\widehat{C}_j^l$ be the  completion time of all the tasks of job $j$ on machine $l$ using Algorithm~1. We now present the following key result.

\begin{theorem}\label{final}
	If all of the jobs are released at time $0$, the weighted  completion time of jobs using S-PC algorithm is upper bounded by $2\left(1+\frac{m-1}{D}\right)$ times the weighted  completion time of jobs under optimal scheduling. Thus, 
	
\begin{equation}
\sum_j w_j \widehat{C}_j\leq 2\left(1+\frac{m-1}{D}\right) \sum_j w_j {C}_j^{OPT}.\end{equation}
	
	For general release times, the weighted  completion time of jobs using S-PC algorithm is upper bounded by $1+2\left(1+\frac{m-1}{D}\right)$ times the weighted  completion time of jobs under optimal scheduling. Thus,  
\begin{equation}
\sum_j w_j \widehat{C}_j \leq \left[1+2\left(1+\frac{m-1}{D}\right)\right]\sum_j w_j {C}_j^{OPT}.
\end{equation}
\end{theorem}

The rest of the section is focused on proving this result. We first note that due to \eqref{lplessopt}, it is enough to show that for zero release times, the following holds for all $k$,  

\begin{equation}
\widehat{C}_{\sigma(k)}\leq 2\left(1+\frac{m-1}{D}\right) {C}_{\sigma(k)}^{LP}. \label{eqshow0rel}
\end{equation}

This, jointly with \eqref{lplessopt}, will prove the desired result. For general release times, it is similarly enough to show for all $k$ that 
\begin{equation}
\widehat{C}_{\sigma(k)}\leq \left[1+2\left(1+\frac{m-1}{D}\right)\right] \label{eqshowgrel} {C}_{\sigma(k)}^{LP}.
\end{equation}

We will first show the result in \eqref{eqshow0rel}, and then show the extensions for \eqref{eqshowgrel}. The result in \eqref{eqshow0rel} will be shown in three steps. The first step evaluates $\widehat{C}_{\sigma(k)}^l$ in terms of $\widehat{C}_{\sigma(k-1)}$. The second step extends this further to write $\widehat{C}_{\sigma(k)}$ in terms of $\widehat{C}_{\sigma(k-1)}$.  The third step uses the result from the second step and performs algebraic manipulations to obtain the result. 

\subsection{Step 1 for the Proof of Theorem \ref{final} for Zero Release Times}

In the first step, we find the time needed to process  job $\sigma(k)$ on machine $l$ once job $\sigma(k-1)$ is completed on every machine. We let $\widehat{C}_{\sigma(0)}=0$. In this subsection, we will show the following result. 

%Based on the greedy algorithm  for assigning tasks on heterogeneous machines given in Algorithm \ref{1job}, we focus on the assignment of the last finished task of job $\sigma(k)$ among all machines. Note that if we assign the last finished task on an arbitrary machine, the maximum expected $\sigma(k)$ processing time of all machines will be extended. By the end of the step, an inequality between the maximum expected processing time of $\sigma(k)$ among all machines and $\mathbb{E}[p_{\sigma(k)}]/\mu$ will be given in cref{max}.

% This result will be proven using three steps. The first step 

%We show the proof of our final result \cref{final} by three steps. 

%In the first step, we analyze the upper bound of expected maximum processing time for each job under  \cref{1job}, an algorithm for scheduling tasks of each job. For every machine, we consider the time needed for processing only job $\sigma(k)$ once job $\sigma(k-1)$ is completed on every machine. Based on the greedy algorithm \cref{1job} for assigning tasks on heterogeneous machines, we focus on the assignment of the last finished task of job $\sigma(k)$ among all machines in practice. Note that if we assign the last finished task on an arbitrary machine, the maximum expected $\sigma(k)$ processing time of all machines will be extended. By the end of the step, an inequality between the maximum expected processing time of $\sigma(k)$ among all machines and $\mathbb{E}[p_{\sigma(k)}]/\mu$ will be given in cref{max}.

%

%$\diamond$ Step 1
\begin{lemma}\label{max}
The difference between the  completion time of all tasks of job $\sigma(k)$ on machine $l$ and the  completion time of all tasks of job $\sigma(k-1)$ is bounded by $\frac{p_{\sigma(k)}}{\mu}(1+\frac{m-1}{D})$. More precisely, we have
\begin{equation}\label{eqmax}
\widehat{C}_{\sigma(k)}^l-\widehat{C}_{\sigma(k-1)}\leq \frac{p_{\sigma(k)}}{\mu}\left(1+\frac{m-1}{D}\right).
\end{equation}
\end{lemma}
%\begin{remark}
	We note that $\frac{p_{\sigma(k)}}{\mu}$ gives the processing time of job $\sigma(k)$ on all machines considered together as a single machine. This result shows an additional factor loss as compared to scheduling on a single machine. 
%\end{remark}

\begin{proof}

%	If we delay the start of  job $\sigma(k)$, the bound on $\widehat{C}_{\sigma(k)}^l-\widehat{C}_{\sigma(k-1)}$ would only be worse and thus we assume that the processing of tasks in job $\sigma(k)$ start after all the tasks in job $\sigma(k-1)$ are complete. 	

%Note that based on the \cref{1job}, $\sigma(k)$ starts when every task of $\sigma(k-1)$ is completed (all machines become idle), $\widehat{C}_{\sigma(k)}-\widehat{C}_{\sigma(k-1)}$ is the maximum processing time of all machines processing only job $\sigma(k)$. For any machine $l$, suppose the completion time for job $\sigma(k)$ is $\widehat{C}_{\sigma(k)}^l$. 
%$$\widehat{C}_{\sigma(k)}-\widehat{C}_{\sigma(k-1)}=\max_{l\in \{1,\cdots, m\}}\{\widehat{C}_{\sigma(k)}^l-\widehat{C}_{\sigma(k-1)}\}.$$

Since the tasks of job $\sigma(k)$ do not have precedence constraints among each other, every task can be started as soon as the previous ones scheduled on a machine are complete. Suppose the task of job $\sigma(k)$ that finishes the last is $t^*$. Let the total  size of all tasks in job $\sigma(k)$ except $t^*$ that are assigned on machine $l$ be denoted by $p_{\sigma(k)}^l$. Then, 
\begin{equation}
p_{\sigma(k)}-p_{\sigma(k),t^*}=\sum_{l=1}^m p_{\sigma(k)}^l.
\end{equation}

%of job $\sigma(k)$ finished last is $t^*$, and the total expected size of all tasks in job $\sigma(k)$ except $t^*$ to be assigned on machine $l$ is $\mathbb{E}[p_{\sigma(k)}^l]$. It is straight forward that $$\mathbb{E}[p_{\sigma(k)}]-\mathbb{E}[p_{\sigma(k),t^*}]=\sum_{l=1}^m \mathbb{E}[p_{\sigma(k)}^l].$$ 

The load for job $\sigma(k)$ on machine $l$ is at most $p_{\sigma(k)}^l+p_{\sigma(k),t^*}$. The  completion time for job $\sigma(k)$ on server $l$ will be highest if task $t^*$ is assigned to this machine. Since the task $t^*$ is assigned to the machine that has the lowest  completion time, the  load for any machine $l$ under the proposed algorithm is upper bounded by $(p_{\sigma(k)}^l+p_{\sigma(k),t^*})/v_l$. Thus we have

%If we assign $t^*$ to an arbitrary machine $l$, the expected completion time on 

%total expected processing time on $l$ will be greater or equal to $\max_{l\in \{1,\cdots, m\}}\mathbb{E}[\widehat{C}_{\sigma(k)}^l-\widehat{C}_{\sigma(k-1)}]$ based on line 4-5 in \cref{1job}. Namely,
\begin{equation}
\max_{l\in\{1,\cdots,m\}}\left(\widehat{C}_{\sigma(k)}^l-\widehat{C}_{\sigma(k-1)}\right)\leq \frac{p_{\sigma(k)}^l+p_{\sigma(k),t^*}}{v_l}, \quad \forall l. \label{refadd}
\end{equation}

 From \eqref{refadd}, we have %Summing the expected load for all machines, and consider the total processing ability during time $\widehat{C}_{\sigma(k-1)}$ to $\widehat{C}_{\sigma(k)}$, we have
\begin{equation}
 v_l \left(\max_{l\in\{1,\cdots,m\}}\widehat{C}_{\sigma(k)}^l-\widehat{C}_{\sigma(k-1)}\right)\leq  {p_{\sigma(k)}^l+p_{\sigma(k),t^*}}, \quad \forall l.
\end{equation}
Adding this over all $l$, we obtain 

\begin{eqnarray}
&&\max_{l\in\{1,\cdots,m\}}\left(\widehat{C}_{\sigma(k)}^l-\widehat{C}_{\sigma(k-1)}\right) \nonumber\\
&\leq& \frac{p_{\sigma(k)}+(m-1)p_{\sigma(k),t^*}}{\mu}\nonumber\\
& \leq& \frac{p_{\sigma(k)}+(m-1)p_{\sigma(k),1}}{\mu}\nonumber\\
& \leq& \frac{p_{\sigma(k)}}{\mu}\left(1+\frac{m-1}{D}\right),
\end{eqnarray}
where $p_{\sigma(k),t^*}\leq p_{\sigma(k),1}$, and $D\le {p_{\sigma(k)}}/{p_{\sigma(k),1}}\le {p_{\sigma(k)}}/{p_{\sigma(k),t^*}}$ by definition.

% or
%\begin{equation}
%\max_{l\in\{1,\cdots,m\}}\mathbb{E}[\widehat{C}_{\sigma(k)}^l-\widehat{C}_{\sigma(k-1)}]\leq \frac{\mathbb{E}[p_{\sigma(k)}]+(m-1)\mathbb{E}[p_{\sigma(k),t^*}]}{\mu}\leq \frac{\mathbb{E}[p_{\sigma(k)}]}{\mu}(1+\frac{m-1}{D}).
%\end{equation}
%where $D={\mathbb{E}[p_{\sigma(k)}]}/{\mathbb{E}[p_{\sigma(k),1}]}$. (Recall that $\mathbb{E}[p_{\sigma(k),1}]=\max_{t\in\{1,\cdots,T_j\}}\mathbb{E}[p_{\sigma(k),t}]\geq \mathbb{E}[p_{\sigma(k),t^*}]$.)
%Now we have completed the proof.
\end{proof}

\subsection{Step 2 for the Proof of Theorem \ref{final} for Zero Release Times}

In the second step, we extend the result in Lemma \ref{max} to get a bound in terms of $\widehat{C}_{\sigma(k)}$ rather than $\widehat{C}_{\sigma(k)}^l$, as given in the following result. %The key issue is that the maximum and the expectation operations are not commutative and thus $\widehat{C}_{\sigma(k)}$ is not the maximum of all $\widehat{C}_{\sigma(k)}^l$. Since we need to understand the random variables of completion times using the algorithm rather than just the means, we denote $A_j^l$ as the random variable corresponding to  the actual completion of job $j$ on machine $l$. Further, $A_j = \max_l A_j^l$ is the random variable corresponding to  the actual completion of job $j$ on all machines. We also have that $\widehat{C}_{\sigma(k)}^l= {\mathbb E}[A_j^l]$, and $\widehat{C}_{\sigma(k)}= {\mathbb E}[A_j]$. 
%We intend to show a bound on the difference between the  completion times of jobs $\sigma(k)$ and $\sigma(k-1)$ as given in the following result. 

\begin{lemma}\label{1jobub}
The difference between the completion times of jobs $\sigma(k)$ and $\sigma(k-1)$ using S-PC algorithm is bounded as follows. 
\begin{equation}
\widehat{C}_{{\sigma(k)}}-\widehat{C}_{{\sigma(k-1)}}\leq \left(1+\frac{m-1}{D}\right)\frac{p_\sigma(k)}{\mu}.
\end{equation}
\end{lemma}
\begin{proof}
	Note that $\widehat{C}_j = \max_l \widehat{C}_j^l$. Thus, 
	
	%Recall that $\widehat{C}_{\sigma(k)}^l$ is the completion time for job $\sigma(k)$ on machine $l$. Base on line 6 in \cref{CMNS}, the processing time for job $\sigma(k)$, $\widehat{C}_{\sigma(k)}-\widehat{C}_{\sigma(k-1)}$ is the maximum of all $\widehat{C}_{\sigma(k)}^l-\widehat{C}_{\sigma(k-1)}$. Namely,
\begin{eqnarray}
	\widehat{C}_{\sigma(k)}-\widehat{C}_{\sigma(k-1)}&=&\max_{l\in\{1,\cdots,m\}}[\widehat{C}_{\sigma(k)}^l-\widehat{C}_{\sigma(k-1)}]\nonumber\\
	&\leq & \left(1+\frac{m-1}{D}\right)\frac{p_\sigma(k)}{\mu},
\end{eqnarray}
	where the inequality follows from \eqref{eqmax} in Lemma \ref{max}.
\end{proof}

%discuss the expected time between completion time of job $\sigma(k)$ and job $\sigma(k-1)$ compared with $\mathbb{E}[p_{\sigma(k)}]/\mu$. Note that the maximum expected processing time of $\sigma(k)$ among all machines and the expected time of the maximum processing time of $\sigma(k)$ among all machines are different since they exchanged expectation and maximum compared to each other. With given coefficient of variation upper bound $\Delta$ for all stochastic tasks involved for job $\sigma(k)$, their relation can be derived based on \cref{sqrt}. The result of step $2$ is shown in \cref{1jobub}.

\subsection{Step 3 for the Proof of Theorem \ref{final} for Zero Release Times}

In this subsection, we will show the result in \eqref{eqshow0rel}. We first show a bound on the  job completion times for the LP-Schedule in the following lemma. 

%To prove the above theorem, we first give a inequality between the total expected processing time of jobs $\sigma(1),\cdots, \sigma(j)$ and the $C_{\sigma(j)}^{LP}$ we get from the LP-Relaxation solution.

\begin{lemma}\label{rel}
Let ${\sigma(1)},\cdots,{\sigma(N)}$ be the schedule of the jobs in S-PC algorithm. Then, 
%For a set of jobs ${\sigma(1)},\cdots,{\sigma(N)}$, if $C_{{\sigma(1)}}^{LP}\leq \widehat{C}_{{\sigma(2)}}^{LP}\leq \cdots \leq \widehat{C}_{{\sigma(N)}}^{LP}$ from the LP-relaxation solution, we have
\begin{equation}
\sum_{k=1}^jp_{\sigma(k)}\leq 2\mu \cdot {C}_{\sigma(j)}^{LP}.
\end{equation}
\end{lemma}
\begin{proof}
\begin{eqnarray*}
C_{\sigma(j)}^{LP}\sum_{k=1}^j {p_{\sigma(k)}}& \geq& \sum_{k=1}^j p_{\sigma(k)}C_{\sigma(k)}^{LP}\\
&\geq& \frac{(\sum_{k=1}^j p_{\sigma(k)})^2}{2\mu} +\sum_{k=1}^j \frac{(p_{\sigma(k)})^2}{2\mu_j}\\
&\geq& \frac{(\sum_{k=1}^j p_{\sigma(k)})^2}{2\mu},
\end{eqnarray*}
where the first inequality is from $C_{{\sigma(1)}}^{LP}\leq C_{{\sigma(2)}}^{LP}\leq \cdots \leq C_{{\sigma(N)}}^{LP}$, and the second inequality is from \eqref{eq2}.
Since every term is non-negative, we further have
\begin{equation}
C_{\sigma(j)}^{LP}\geq \frac{\sum_{k=1}^j p_{\sigma(k)}}{2\mu}.
\end{equation}
\end{proof}
%Note that if we only process tasks in job $j$ on all machines, the optimal processing time at least $\mathbb{E}[p_j]/\mu$.
%For any job $j$ in the list of jobs by ascending order of $C_j^{LP}$, we denote $B_j$ as the set of jobs before (containing) $j$. We consider a fixed scheduling $p$ constructed by \cref{CMNS}. 

We will now describe the steps to prove \eqref{eqshow0rel}, which proves  Theorem \ref{final} for zero release times. 

\begin{proof}[Proof of \eqref{eqshow0rel}]
%Recall that \eqref{eq1} to \eqref{eq4} are necessary conditions for optimal schedule, 
%\begin{equation}\label{3th}
%\mathbb{E}[C_{\sigma(k)}^{OPT}]\geq C_{\sigma(k)}^{LP}\geq \frac{\sum_{k=1}^j \mathbb{E}[p_{\sigma(k)}]}{2\mu},\quad k\in\{1,\cdots,N\}.
%\end{equation}

From Lemma \ref{1jobub}, we have %note that we regard $\widehat{C}_{\sigma(0)}$ as $0$:
\begin{equation}
\widehat{C}_{{\sigma(k)}}-\widehat{C}_{{\sigma(k-1)}}\leq \left(1+\frac{m-1}{D}\right)\frac{p_{\sigma(k)}}{\mu}. 
\end{equation}

Summing this over $k$ from $1$ to $j$, we have

\begin{equation}
\widehat{C}_{\sigma(j)}  \leq \left(1+\frac{m-1}{D}\right)\frac{\sum_{k=1}^{j}p_{\sigma(k)}}{\mu}.
\end{equation}

Further, using Lemma \ref{rel}, we have

\begin{equation}
\widehat{C}_{\sigma(j)} \leq 2\left(1+\frac{m-1}{D}\right) C_{\sigma(j)}^{LP}. 
\end{equation}
\end{proof}

\subsection{Extension to General Release Times}

%where the second inequality comes from adding $\mathbb{E}[\widehat{C}_{{\sigma(k)}}-\widehat{C}_{{\sigma(k-1)}}]$ up, and the third inequality follows \eqref{3th}. Now we have proved the inequality when there is no release time.

When release times are not zero, \eqref{eq3} indicates that for any job $j$, $ C_{j}^{LP}\geq r_j$. Therefore, if the job $\sigma(k)$ starts after $r_{\sigma(k)}$ in addition to the completion of previous jobs, then 

\begin{equation}
\widehat{C}_{\sigma(k)} \leq r_{\sigma(k)}+ 2\left(1+\frac{m-1}{D}\right) C_{\sigma(k)}^{LP}. 
\end{equation}
Using $ C_{j}^{LP}\geq r_j$, we have the result as in  \eqref{eqshowgrel}, which proves  Theorem \ref{final} for general release times.

%\begin{equation*}
%\mathbb{E}[\widehat{C}_{\sigma(k)}] \leq [1+2(1+\sqrt{m}\Delta)(1+\frac{m-1}{D})]\mathbb{E}[C_{\sigma(k)}^{OPT}].
%\end{equation*}
%Now, we have completed the proof.

%\input{approx_results}
%\vspace{-.1in}

%\begin{figure*}[]
%	\begin{minipage}{0.5\textwidth}
%		\begin{center}
%			\includegraphics[height=1.65in, width=\textwidth]{./figures/detailsO.pdf}%
%			\caption{Default Hadoop's progress score vs. normalized (actual) execution time. }
%			\label{fig:def}%
%		\end{center}
%	\end{minipage}
%	\hspace{0.1cm}
%	\begin{minipage}{0.5\textwidth}
%		\begin{center}
%			\includegraphics[height=1.65in, width=\textwidth]{./figures/details.pdf}%
			%\vspace*{-.5cm}
%			\caption{Progress score of proposed estimator vs. normalized (actual) execution time. }
%			\label{fig:imp}%
%		\end{center}
%	\end{minipage}
%\end{figure*}

\section{Implementation } \label{sec:implementation}  

%\subsection{Implementation}
We implement our proposed scheduler and evaluate it in Hadoop, whose map and reduce tasks satisfy a complete bipartite dependence graph. Our implementation consists of three key modules: a {\em job scheduler} that implements Algorithm~1 to determine the scheduling order of different jobs, a {\em task scheduler} that is responsible for scheduling map and reduce tasks on different machines, and an {\em execution database} that stores statistics of previously executed jobs/tasks for estimating task completion times. A key feature of our implementation is its ability to adapt to system dynamics and task interference at runtime, which introduce additional sources of processing-time uncertainty. In particular, S-PC's task scheduler not only computes an optimal schedule of map and reduce tasks according to Algorithm~1, but also has the ability to re-optimize the schedule on the fly based on available task progress and renewed completion time estimates, as it continuously monitors progress of map/reduce tasks through collaboration with Hadoop's Application Master module and updates estimates of the remaining completion time. Further, to cope with potential desynchronization and disconnection issues in Hadoop (which may cause task execution to deviate from the optimal schedule), we also modify the Application Master (AM) and Resource Manager (RM) in Hadoop, which work collectively with task scheduler to ensure the execution (and resumption) of tasks in the desired order. 

\begin{figure}[thbp]
	%\vspace{-.05in}
	\centering
	\includegraphics[height=2in, width=0.4\textwidth]{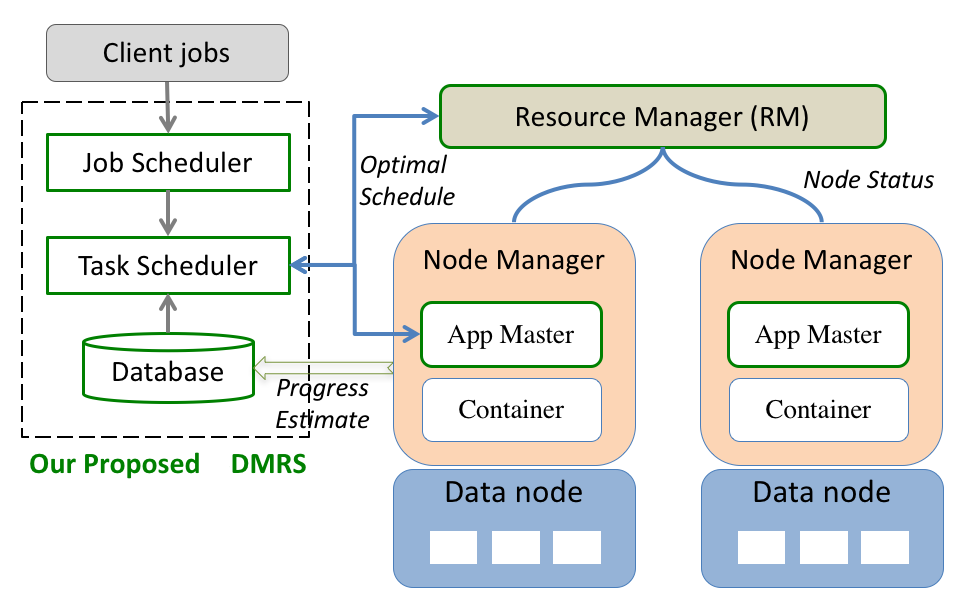}
	\caption{\small System diagram of our proposed scheduler implementation.}
	\label{fig:impl}
	%\vspace{-.45in}
\end{figure}

More precisely, our scheduler works as follows. First, the job scheduler loads necessary job parameters and queries the execution database for estimated machine speeds (speed of machine $l$ is $v_l$), to formulate and solve the LP-Schedule problem. The optimal job schedule is input to the task scheduler to find the schedule and placement of every map and reduce task according to Algorithm~1. Next, based on the task schedule and placement, the RM assigns a queue to each machine to store all map and reduce tasks that are scheduled to run on it. Tasks in each machine $l$'s queue are then processed in a FIFO manner, guaranteeing the execution of jobs/tasks under our proposed algorithm. In particular, each task is given a unique ID. When resources become available on machine $l$, the RM launches a container and associates it with the head-of-line task. The container and task-ID pair are sent to the AM for launching the desired task.

\begin{figure}[thbp]
	%\vspace{-.05in}
	\centering
	\includegraphics[width=0.49\textwidth]{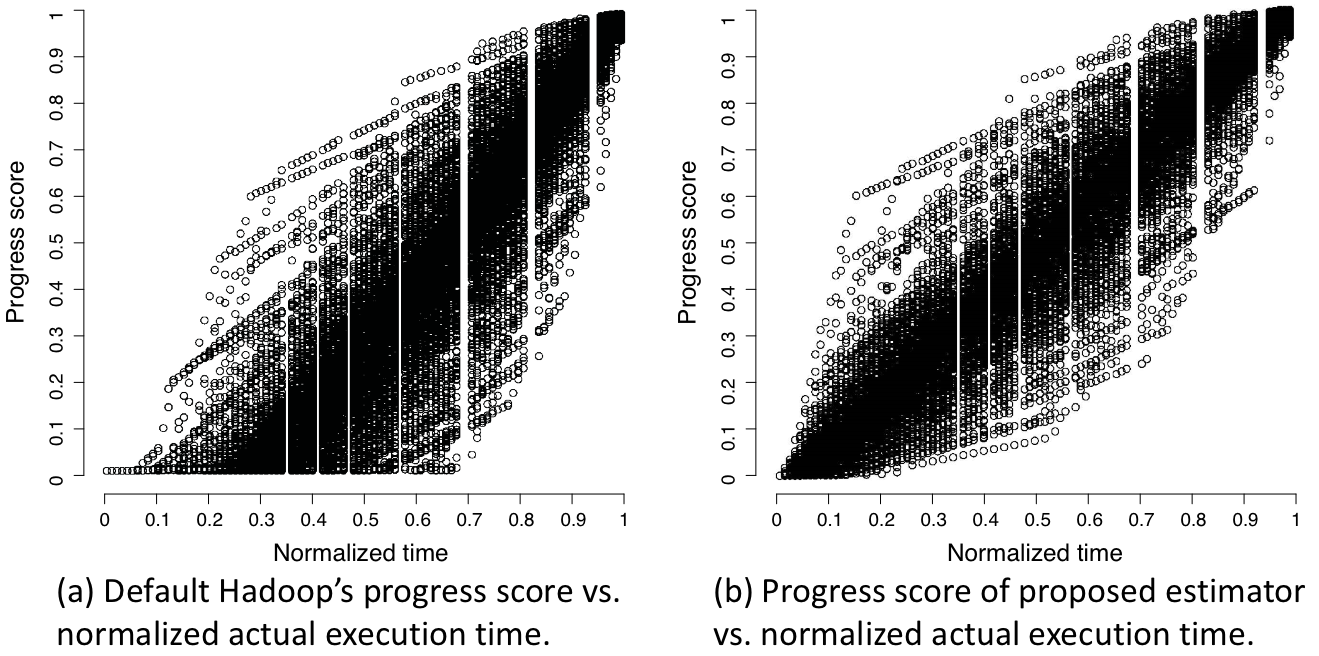}
	\caption{\small The proposed estimator significantly improves progress estimates: (a) Default Hadoop's progress score and (b) Proposed estimator's progress score vs. normalized actual execution time.}
	\label{fig:est}
	%\vspace{-.45in}
\end{figure}

Next, before launching each (map or reduce) task $t$, the task schedule estimates the completion time $t_l$ of all jobs/tasks scheduled before $t$ on each machine $l=1,2,\ldots,m$. The time $t_l$ is obtained by combining known task completion times (which are available from execution database) and estimating the remaining times of active tasks (which are calculated by each AM using the remaining data size divided by machine speed). In particular, we continuously monitor task/job progress through AMs, refines the estimate of completion time, and if necessary, re-optimizes task schedules on the fly. By default, 
Hadoop reports progress scores for each task and provides estimated task execution time, derived as the time elapsed since task launching divided by the current reported progress score. In Figure.~\ref{fig:est}, we run 1000 tasks on Default Hadoop, and plot the progress score from Hadoop Reporter every 3 seconds against the (normalized) actual task completion times. It can be seen that the default progress score and actual task completion times do not follow a linear relationship, leading to high inaccurate time estimates. To mitigate this issue, we recognize that task completion time estimation error is mainly caused by Hadoop's assumption that a task starts running right after it is launched. However, due to highly contended resource-sharing environments, JVM startup time is significant, and we need to take into consideration the time to launch a JVM when estimating task progress and completion time. In particular, we calculate the time for launching JVM by finding the difference between first progress report time ($t_{\rm FP}$) and launching time ($t_{\rm lau}$). Therefore, the new estimated completion time is given by
\begin{equation}\label{t_of}
t_{\rm ect} = t_{\rm lau} + (t_{\rm FP}-t_{\rm lau}) + ({t_{\rm now}-t_{\rm FP}}) \/ ({CP-FP}),
\end{equation}
where $\frac{t_{\rm now}-t_{\rm FP}}{CP-FP}$ is time for processing workload, and $FP$ and $CP$ are first reported progress value and current reported progress value, respectively.  Figure.~\ref{fig:est} shows that our proposed estimator can eliminate the bias and provide more reliable estimation of task progress and required completion time, further boosting the performance of our scheduling algorithm.

%\begin{figure}[]
%\begin{minipage}{0.4\textwidth}
%\begin{center}
%		\includegraphics[width=0.8\textwidth]{./figures/detailsO.pdf}%
%		\caption{Default Hadoop's progress score vs. normalized (actual) execution time. }
%				\label{fig:def}%
%\end{center}
%\end{minipage}
%\hspace{0.1cm}
%\begin{minipage}{0.4\textwidth}
%\begin{center}
%		\includegraphics[width=0.8\textwidth]{./figures/details.pdf}%
		   %\vspace*{-.5cm}
%		\caption{Progress score of proposed estimator vs. normalized (actual) execution time. }
%	\label{fig:imp}%
%\end{center}
%\end{minipage}
%\vspace{-.2in}
%\end{figure}

Then, the optimization in Algorithm~1 is repeated at runtime to find the optimal machine $l^*$ for task $t$. A new optimization of all remaining tasks by the task scheduler is triggered if $l^*$ is different from the previous solution. This makes our S-PC schedule robust to any possible execution uncertainty and estimation errors. We also implement additional features in both AM and RM to make them fault tolerant. The container and task-ID pairs are duplicated at each AM in advance (after an optimal schedule is computed by the job and task schedulers). If RM accidentally sends an incorrect container that is intended for application (\eg due to lack of synchronization), AM will detect such inconsistency and immediately release the container back to RM. Further, a mechanism to handle  occasional disconnection is implemented in both AM and RM, allowing them to buffer current containers/tasks and attempt reconnection.
%\vspace{-.1in}

\section{Evaluation} \label{sec:evaluation}  

%We evaluate the proposed S-PC on a heterogeneous cluster and with respect to real-world datasets.

\subsection{Evaluations of S-PC.}

\begin{figure*}[!t]
	\centering
	
	\subfigure[]{%
		\includegraphics[height=1.55in,width=0.33\textwidth]{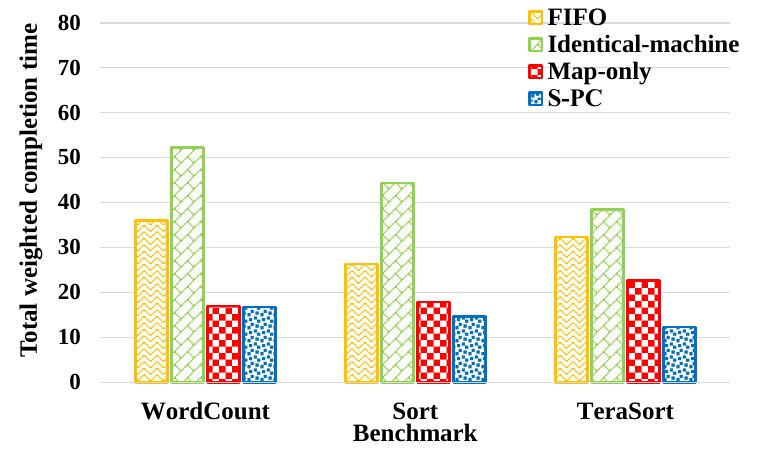}%
		% \vspace*{-.2cm}
		\label{fig:x1}%
	}%
	%\hfill
	~
	\centering
	\subfigure[]{%
		\includegraphics[height=1.55in, width=0.33\textwidth]{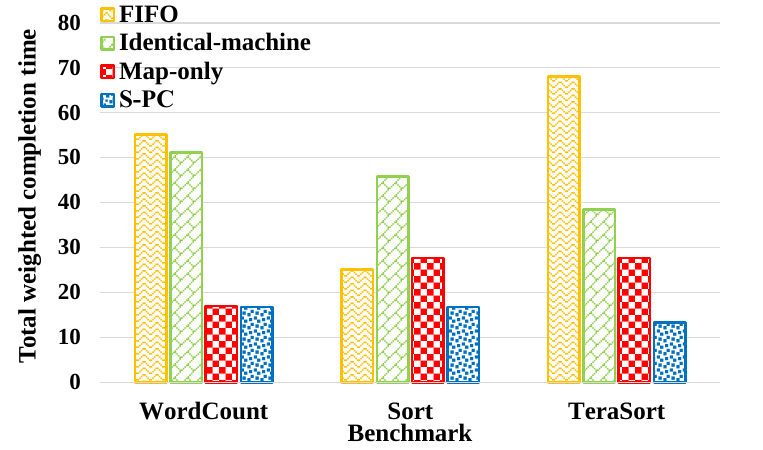}%
		%    \vspace*{-.2cm}
		\label{fig:x2}%
	}%
	%\hfill
	~
	% \centering
	\subfigure[]{%
		\includegraphics[height=1.55in,width=0.33\textwidth]{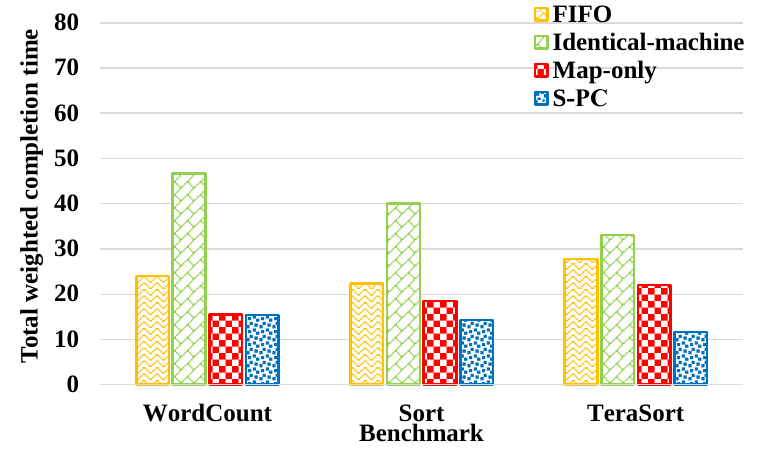}%
		%    \vspace*{-.2cm}
		\label{fig:x3}%
	}%
	%\vspace{-.15in}
	\caption{\small (a) Total weighted completion time of jobs with identical task sizes and workloads. (b) Total weighted completion time of jobs with different task sizes and workloads. (c) Total weighted completion time of jobs with different task sizes and workloads.}
	\label{fig2}
		%\vspace{-.3in}
\end{figure*}

We evaluate our scheduler on a Hadoop cluster with three benchmarks, viz., WordCount, Sort, and TeraSort. We compare S-PC with FIFO, Identical-machine, and Map-only schedulers. FIFO is provided by Hadoop, and FIFO schedules jobs based on the jobs' releasing order. Within a job, FIFO schedules a task to the first available machine, and reduce tasks are scheduled after the map tasks of the job are completed. Identical-machine assumes all machines are identical, and applies Algorithm~1 to schedule jobs and tasks. Map-only considers the map phase is the most critical, and employs Algorithm~1 to schedule jobs and tasks without considering the reduce phase.

{\noindent \bf Experiment setup:} We set up a heterogeneous cluster. The cluster contains $12$ (virtual) machines, and each machine consists of a physical core and 8GB memory. Each machine can process one task at a time. In the cluster, machines are connected to a gigabit ethernet switch and the link bandwidth is 1Gbps. The heterogeneous cluster contains two types of machines, fast machines and slow machines. The processing speed ratio between a fast machine and a slow machine is $8$. We evaluate our scheduler by using three benchmarks -- WordCount, Sort, and TeraSort. WordCount is a CPU-bound application, and Sort is an I/O-bound application. TeraSort is CPU-bound for map phase, and I/O bound for reduce phase. We download workload for WordCount from Wikipedia, and generate workloads for Sort and TeraSort by using RandomWriter and TeraGen applications provided by Default Hadoop. The number of reduce tasks per job is set based on workload of the reduce phase. We set the number of reduce tasks per job in WordCount to be 1, and in Sort and TeraSort to be 4. All jobs are associated with weights, which are uniformly distributed between 1 and 5. Also, all jobs are partitioned into two releasing groups, and each group contains the same number of jobs. The releasing time interval between the two groups is $60$sec. %The completion time of a job is measured by the hour.

%\subsection{Experimental results}
{\noindent \bf Experiment results:} In the first set of experiments, each experiment contains 20 jobs, and the workload of a job is 1GB. The task sizes of all jobs are the same, and equal 64MB. Figure~\ref{fig:x1} shows that our scheduler outperforms FIFO, Identical-machine, and Map-only by up to $62\%$, $68\%$, and $45\%$, respectively. Identical-machine has the largest total weighted completion time (TWCT). The reason is that it distributes the same number of tasks to each machine. A job's completion time is dominated by the tasks' completion time running on slow machines. Also, Identical-machine results in a large amount of cluster resources being wasted, since fast machines need to wait for slow machines to finish their tasks. FIFO schedules jobs based on the jobs' release order. Jobs with high weights (time-sensitive jobs) cannot be scheduled first, so time-sensitive jobs cannot be completed in time. For task scheduling, FIFO does not consider the heterogeneous cluster environment, and tasks are scheduled to the first available container. Such a task scheduling scheme can increase the completion time of tasks, since a container which becomes available later might be launched on a fast machine and be able to complete a task faster. Also, FIFO might schedule reduce tasks soon after the job is scheduled, and before the last map task is scheduled. Even though such scheme leaves more time for reduce tasks to fetch data from map tasks' outputs, given the large available network bandwidth nowadays, reduce tasks only need a little time for fetching data from all map tasks' outputs. Map-only schedules jobs and tasks without considering the reduce phase. Under benchmarks with light workloads in the reduce phase, \eg under WordCount benchmark, Map-only can achieve comparable performance as our scheduler. However, if the workload of the reduce phase is comparable with the map phase's, \eg under Sort, scheduling jobs without considering the workload of reduce phases and scheduling reduce tasks to random machines result in performance degradation and increase in TWCT. Furthermore, under a benchmark with heavy workload in the reduce phase, \eg TeraSort, TWCT is dominated by the completion time of reduce tasks, and Map-only increases TWCT by $85\%$, compared with our scheduler.

%\begin{figure}[!t]
%	\centering
%	\includegraphics[width=0.46\textwidth]{./figures/new4.pdf}
%	\caption{\small Total weighted completion time under different ratios of the number of elephant jobs to total jobs.}
%	\label{fig:y}
%\end{figure}

\begin{figure*}[!thp]
\begin{minipage}[ht]{0.31\textwidth}	
	%\subfigure[Comparing storage capacity] {
	\vspace{0.05in}
		\includegraphics[width=\textwidth]{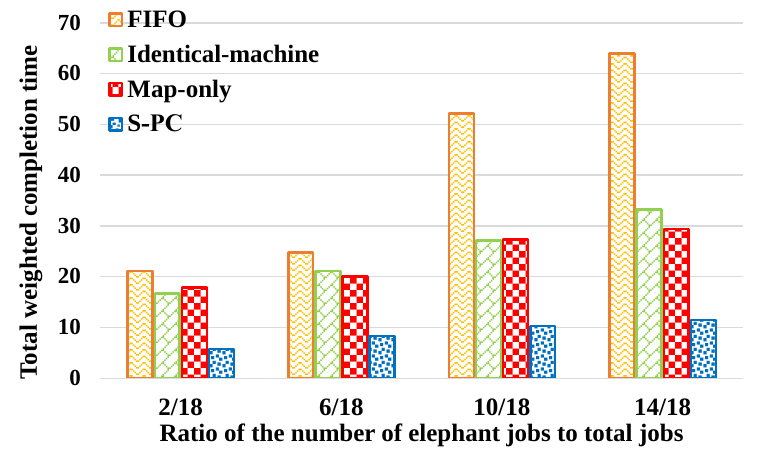}
		\vspace{-0.2in}
			\label{fig:y1}
		\caption{Total weighted completion time under different ratios of the number of elephant jobs to total jobs..}
\end{minipage}
\hspace{0.2cm}
\begin{minipage}[ht]{0.62\textwidth}
		\subfigure[Comparing weighted completion time] {
		\includegraphics[width=0.48\textwidth]{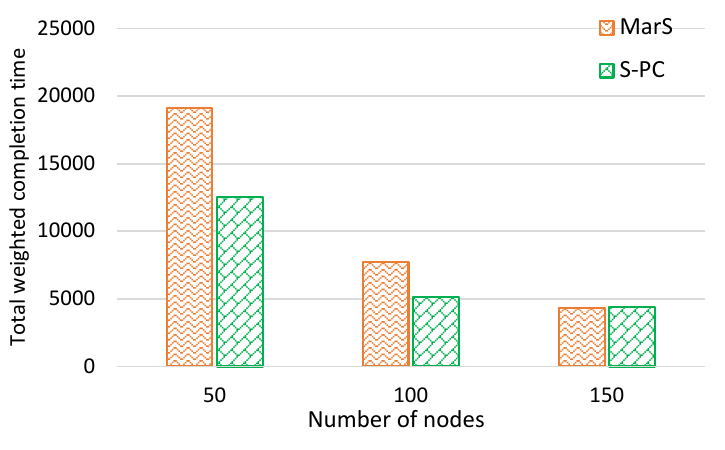} }
		\label{fig:sim10}		
%	\end{minipage}
%	\hspace{0.2cm}
%\begin{minipage}[ht]{0.31\textwidth}	
	\subfigure[Accumulative completion time over $t$] {
		\includegraphics[width=0.48\textwidth]{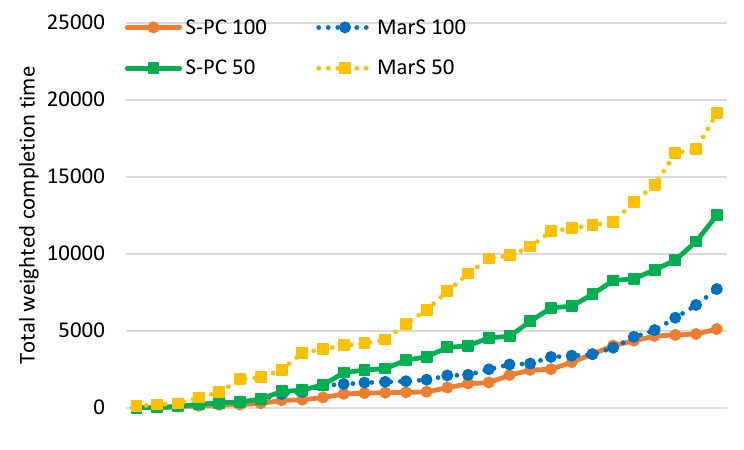} }
		\label{fig:sim20}	
	\caption{Comparing S-PC and MarS via Google-trace simulation and online re-optimization.}
\end{minipage}  
\end{figure*}

Figure~\ref{fig:x2} shows the results of employing jobs with different task sizes and different amount of workloads. In the set of experiments, whose results are shown in Figure~\ref{fig:x2}, each experiment contains 20 jobs. Of these 20 jobs, 12 jobs need to process 1GB data each, and the task size is 64MB; 4 jobs need to process 0.5GB each, with a task size of 32MB; and the remaining 4 jobs need to process 2GB, with a task size of 128MB. Figure~\ref{fig:x2} shows that S-PC outperforms FIFO, Identical-machine, and Map-only by up to $80\%$, $65\%$, and $52\%$, respectively. Under WordCount and TeraSort, the TWCT of FIFO is much larger than TWCT of other schedulers in Figure~\ref{fig:x2} and TWCT of FIFO in Figure~\ref{fig:x1}. This is because several jobs schedule reduce tasks soon after the jobs are scheduled; those reduce tasks occupy all fast machines, and since they cannot start to process data until all map tasks complete, all map tasks end up being scheduled on slow machines. Even though jobs' completion time of FIFO has large variation, based on our results, our scheduler can outperform FIFO by at least $36\%$. Also, introducing jobs with large workloads ({\em large jobs}) increases the TWCT of FIFO, since it does not consider jobs' and tasks' workloads in job scheduling. Jobs with small workloads ({\em small jobs}) might be scheduled after large jobs, and this makes small jobs suffer from the starvation problem. 

We further increase the number of large jobs and small jobs, and set the number of jobs in each experiment to be 18 to make the total workload of jobs in each experiment be roughly the same as the first two sets of experiments'. Of the 18 jobs, 6 jobs need to process 1GB data each, with a task size of 64MB; another 6 jobs need to process 0.5GB each, with a task size of 32MB; and the remaining 6 jobs need to process 2GB each, with a task size of 128MB. Figure~\ref{fig:x3} shows that as the number of large jobs increases,  our scheduler has low TWCT, since small jobs with large weights do not suffer from the starvation problem. 

%\ref{fig:y1}
In the final experiment, we fix the number of TeraSort jobs to be 18, and change the number of elephant jobs. We set the task size to be 64MB. An elephant job needs to process 2GB data, and a mice job needs to process 0.5GB data. Figure~5 shows that our scheduler outperforms FIFO, Identical-machine, and Map-only by up to $82\%$, $66\%$, and $61\%$, respectively. As the number of elephant jobs increases, mice jobs with large weights might be scheduled after more elephant jobs, and this results in long waiting times for mice jobs, causing large increase in TWCT. Also, as the number of elephant jobs increases, the total workload increases, and the long time occupied on fast machines by reduce tasks before all map tasks finish increases TWCT greatly, since more map tasks have to process data on slow machines. By comparing TWCT of Identical-machine, Map-only, and our scheduler, we observe that as the number of elephant jobs increases, the TWCT increases for Identical-machine and Map-only are much larger than for our scheduler's. For Identical-machine, increasing the number of elephant jobs means the difference of the amount of time used to complete all assigned tasks on fast machines and on slow machines increases. Without considering the scheduling of reduce tasks, as the number of elephant jobs increases, more workloads of reduce tasks are assigned to slow machines, and this results in a large increase in TWCT.

\begin{figure}[thp!]
%	\noindent\begin{minipage}{0.48\linewidth}
		\centering
		\includegraphics[width=0.48\textwidth]{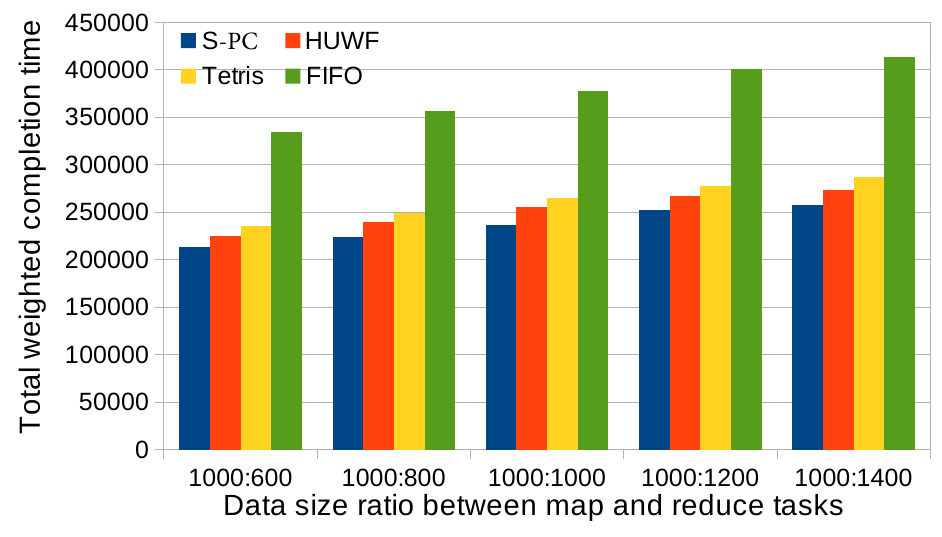}
		\scalebox{0.74}{
			\begin{tabular}{|c|c|c|c|c|c|}
				\hline
				& 1000:600 & 1000:800 & 1000:1000 & 1000:1200 & 1000:1400 \\ \hline
				S-PC 	& 212997   & 224317	  & 237050    & 251986    & 257677	  \\ \hline
				HUWF	& 224689   & 239676	  & 255599    & 267546	  & 273419	  \\ \hline
				Tetris	& 235362   & 249202   & 265137    & 277193	  & 286683	  \\ \hline
				FIFO    & 334947   & 356623   & 377699    & 400461    & 413924    \\ \hline
		\end{tabular}}
		\captionof{figure}{\small Comparisons of S-PC, HUWF, Tetris, and FIFO in terms of total weighted job completion time with different data size ratios between map and reduce tasks.}
		\label{fig:sim1}
		\vspace{.15in}
	\end{figure} 
\begin{figure}[thp!]
		\centering
		\includegraphics[width=0.48\textwidth]{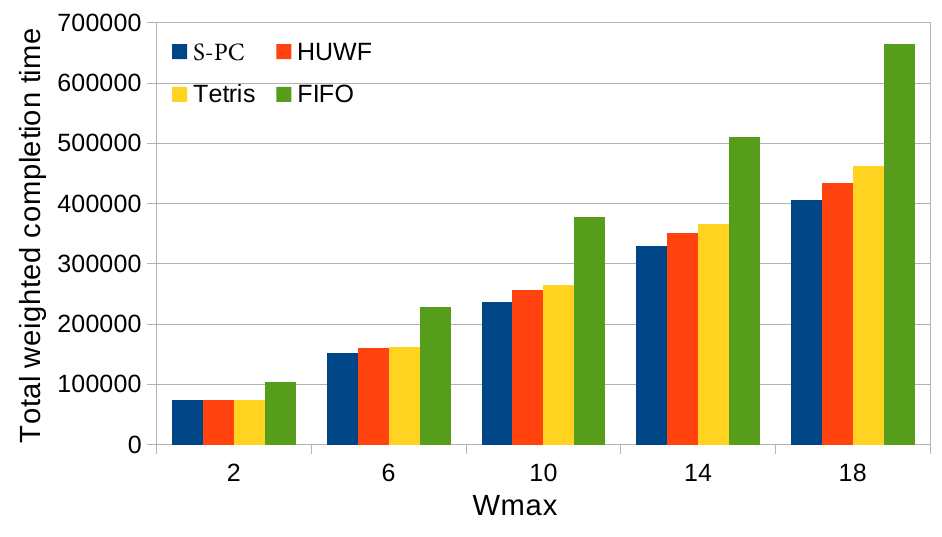}
		\scalebox{0.9}{
			\begin{tabular}{|c|c|c|c|c|c|}
				\hline
				& 2 & 6 & 10 & 14 & 18 \\ \hline
				S-PC 	& 74446   & 152420	  & 237050    & 328727    & 405319	  \\ \hline
				HUWF	& 73295   & 159819	  & 255599    & 350878	  & 434418	  \\ \hline
				Tetris	& 73333   & 162568    & 265137    & 366656	  & 461547	  \\ \hline
				FIFO    & 103014  & 228727    & 377699    & 511059    & 664199    \\ \hline
		\end{tabular}}
		\captionof{figure}{\small Comparisons of S-PC, HUWF, Tetris, and FIFO in terms of total weighted job completion time with different $W_{max}$.}
		\label{fig:sim2}
%	\end{minipage} 
	\vspace{-.15in}
\end{figure}

\if 0
\begin{figure*}[thp!]
\noindent\begin{minipage}{0.48\linewidth}
	\centering
	\includegraphics[width=0.98\textwidth]{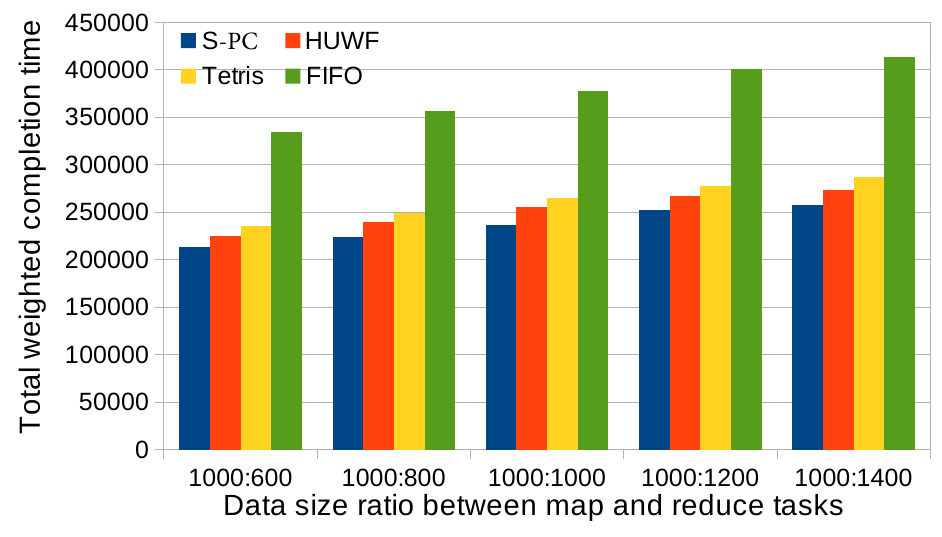}
	\scalebox{0.74}{
	\begin{tabular}{|c|c|c|c|c|c|}
	\hline
	& 1000:600 & 1000:800 & 1000:1000 & 1000:1200 & 1000:1400 \\ \hline
	S-PC 	& 212997   & 224317	  & 237050    & 251986    & 257677	  \\ \hline
	HUWF	& 224689   & 239676	  & 255599    & 267546	  & 273419	  \\ \hline
	Tetris	& 235362   & 249202   & 265137    & 277193	  & 286683	  \\ \hline
	FIFO    & 334947   & 356623   & 377699    & 400461    & 413924    \\ \hline
	\end{tabular}}
	\captionof{figure}{\small Comparisons of S-PC, HUWF, Tetris, and FIFO in terms of total weighted job completion time with different data size ratios between map and reduce tasks.}
	\label{fig:sim1}
	\vspace{.15in}
\end{minipage} 
\hspace{0.2cm}
\noindent\begin{minipage}{0.48\linewidth}
	\centering
	\includegraphics[width=0.98\textwidth]{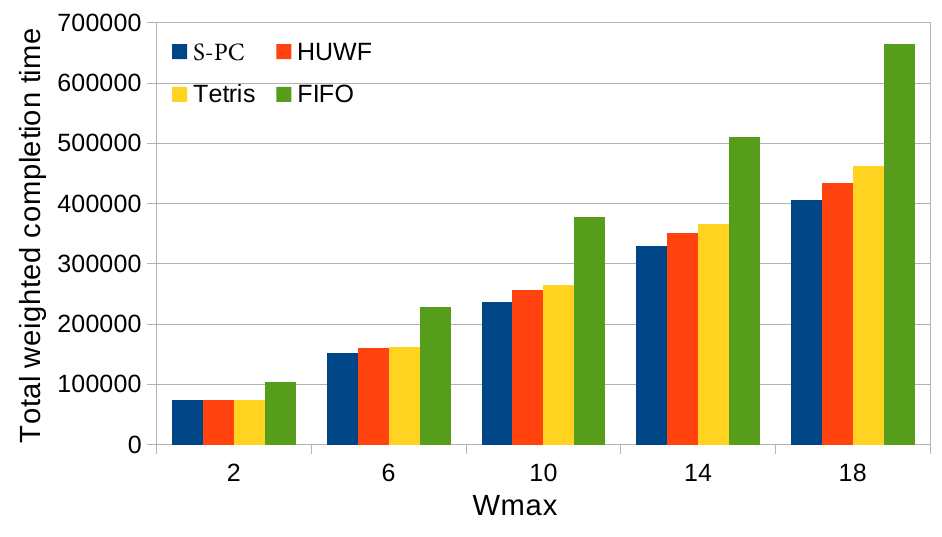}
	\scalebox{0.9}{
		\begin{tabular}{|c|c|c|c|c|c|}
			\hline
			& 2 & 6 & 10 & 14 & 18 \\ \hline
			S-PC 	& 74446   & 152420	  & 237050    & 328727    & 405319	  \\ \hline
			HUWF	& 73295   & 159819	  & 255599    & 350878	  & 434418	  \\ \hline
			Tetris	& 73333   & 162568    & 265137    & 366656	  & 461547	  \\ \hline
			FIFO    & 103014  & 228727    & 377699    & 511059    & 664199    \\ \hline
	\end{tabular}}
	\captionof{figure}{\small Comparisons of S-PC, HUWF, Tetris, and FIFO in terms of total weighted job completion time with different $W_{max}$.}
	\label{fig:sim2}
\end{minipage} 
	\vspace{-.15in}
\end{figure*}
\fi 
	%\vspace{-.2in}
\subsection{Large-scale simulations}\label{sec:simulation}  

{\noindent \bf Evaluation of online re-optimization.}  We conduct a large-scale, trace-driven simulation (using the Google trace~\cite{GoogleTrace}) to evaluate S-PC's ability to re-optimize job schedules on the fly. More precisely, upon each job arrival or departure, S-PC is employed to re-optimize and update the schedules of all existing jobs in the system. For each job, we extract the arrival time, the number of associated tasks, and the workload of each task from Google trace~\cite{GoogleTrace}. Each job is assigned a weight uniformly distributed between 1 and 10, and its tasks are randomly partitioned into map and reduce phases (with $60\%$ and $40\%$ probabilities, respectively). We simulate a  system with $m=50,100,150$ machines, each with a different speed-up ranging from $1$ to $3$. Figure~6(a) shows the total weighted completion time of all jobs and compares S-PC with an existing online scheduler, MarS~\cite{yuan2014joint}, which considers joint map and reduce scheduling for the special case with identical machine speeds. We observe that S-PC outperforms MarS by up to $34\%$, when the system load is high (i.e., a large number of tasks to be processed by each machine as $m=50$). This is because S-PC is able to jointly schedule map and reduce tasks of heterogeneous jobs with respect to the precedence constraints. Furthermore, the evolution of accumulative weighted completion time (over time $t$) is shown in Figure~6(b), for different numbers of machines. We note that as the accumulative weighted completion time of MarS grows rapidly with more jobs arriving over time, the benefit of S-PC becomes more substantial with online re-optimization.

{\noindent \bf Evaluation of scalability.} In each evaluation, we simulate 100 jobs, with multiple rounds of dependent tasks, whose numbers are generated uniformly between 1 and 50.
%and the number of map and reduce tasks in a job are  In a job, the number of reduce tasks is no more than map tasks. 
%We set default data size of a task to be 1000 MB. 
Weights of jobs are generated uniformly between 1 and $W_{max}$, and the default value of $W_{max}$ is 10. Jobs are scheduled in a 100-machine cluster. Data processing speed of machines follows Gaussian distribution with mean=50 MB/sec and standard deviation=10. A machine can process one task at a time. Each data point in the following figures is the average value over 20 evaluations. Since existing scheduling algorithms fail to optimize under such dependence, we compare our scheduler with a few heuristics: High Unit Weight First (HUWF), Tetris~\cite{grandl2015multi}, and FIFO algorithms. In {HUWF}, Unit Weight (UW) of a job equals job's weight divided by job execution time. Jobs are sorted based on UW in descending order, and map and reduce tasks are scheduled one by one. A task is assigned to a machine that produces the earliest completion time. In Tetris, resource usage score of a job equals the number of tasks (machines) multiplied by job execution time. Jobs are sorted based on resource usage scores, and map and reduce tasks are scheduled one by one. A task is assigned to a machine that produces the earliest completion time. Finally, FIFO sorts jobs and schedules tasks one by one to the first available machine.

%\begin{itemize}
%	\item {HUWF:} Unit weight (UW) of a job equals job's weight divided by job execution time, which equals  $(p_j^M + p_j^R)/v_1$. Jobs are sorted based on UW in descending order, and map and reduce tasks are scheduled one by one. A task is assigned to a machine that produces the earliest completion time.
%	\item {Tetris:} Resource usage score of a job equals the number of tasks (machines) multiplying by job execution time, which equals  $(p_j^M + p_j^R)/v_1$. Jobs are sorted based on resource usage scores, and map and reduce tasks are scheduled one by one. A task is assigned to a machine that produces the earliest completion time.
%	\item {FIFO:} FIFO sorts jobs based on job Ids, and then schedules map and reduce tasks one by one. A task is assigned to the first available machine.
%\end{itemize}

Figure~\ref{fig:sim1} compares S-PC with HUWF, Tetris and FIFO, in terms of total weighted completion time, by changing data size ratio between map and reduce tasks. As data size of a reduce task increases from 600 MB to 1400 MB, the total weighted job completion time of S-PC, HUWF, Tetris and FIFO increases by 44680, 48730, 51321, and 78977 ({20.98\%, 21.69\%, 21.8\%, and 23.58\%}), respectively. The results show that by considering precedence constraints between map and reduce tasks in job scheduling, S-PC has smaller increase in total weighted completion time as data sizes of reduce tasks increase.

%\begin{figure}[!t]
%	\centering
%	\includegraphics[width=0.46\textwidth]{./figures/sim2.pdf}
%	\caption{Comparisons of S-PC, HUWF, Tetris, and FIFO in terms of weighted job completion time with different maximum job weights.}
%	\label{fig:sim2}
%\end{figure}

Figure~\ref{fig:sim2} compares our proposed scheduler with HUWF, Tetris and FIFO, in terms of TWCT, by changing maximum job weights. When job weights are small, TWCT of our scheduler, HUWF, and Tetris are almost the same. As maximum job weight increases from 2 to 18, the weighted job completion time of our scheduler, HUWF, Tetris and FIFO increases by 330873, 361123, 388214 and 561185, respectively. Because Tetris and FIFO do not consider weight in job scheduling, as job weight increases, Tetris and FIFO have larger increase in TWCT than our scheduler and HUWF. Results also show that scheduling jobs based on UW is not efficient. To achieve smaller TWCT, we also need to consider precedence between different phases and different machine processing speeds.
%, i.e., Constraints \eqref{const2}, \eqref{const3}, \eqref{const4}, and \eqref{const5}.

\subsection{Stochastic Task Execution Times} 

\begin{figure}[!t]
	\centering
	\includegraphics[width=0.42\textwidth]{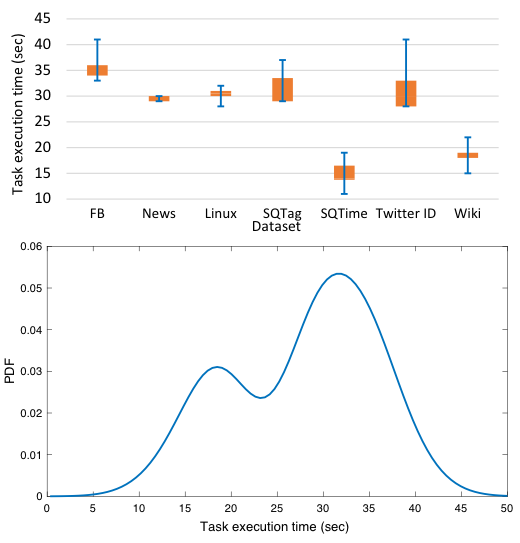}
\caption{Empirical distribution of map task processing time for real-world datasets (using {\em WordMean} and 128MB split size): (top) box-and-whisker plot for each dataset, and (bottom) aggregate {\em cdf} of task processing time.}		\label{fig:inputData}
	\vspace{-.15in}
\end{figure}

We now evaluate the proposed S-PC algorithm under stochastic task execution times.  Although tasks are often designated to process fixed and equal-size data splits, their processing times can still vary significantly in practice, even if machines are assumed to have deterministic speed. For example, programs like {\em WordCount} and {\em WordMean}  need to iteratively store and process unique words from text files. The number of such operations (and thus processing time) required by each individual task depends heavily on the text content. Similarly, the processing times of {\em Sort} and {\em K-mean} are highly dependent on initial data distribution. Figure~\ref{fig:inputData} shows the distribution of map-task processing times on real-world datasets (including Facebook \cite{facebook}, Twitter \cite{twitter}, Linux source files \cite{linux}, News \cite{cnn}, Stack Overflow \cite{stack}, and Wikipedia \cite{wiki}) with a fixed split size of 128MB. Not only do we observe high processing-time uncertainty among different datasets (with {a mean of 28 seconds and a standard deviation of 7 seconds, or 25\% of the mean}, in the aggregate distribution), but there is also substantial randomness when processing the same dataset. 

We show that applying the S-PC algorithm with respect to mean task size $\bar{p}_{j,t}=\mathbb{E}[{p}_{j,t}]$, we can obtain a robust task schedule significantly reducing the total weighted completion time. The S-PC algorithm is evaluated against four baselines: FIFO, Identical-machine, Map-only, LATE~\cite{LATE2008} and MarS~\cite{yuan2014joint} schedulers, with respect to stochastic task execution times and three benchmarks, {\em viz.}, WordCount, WordMean, and Sort. In particular, LATE is a popular scheduler for speculatively scheduling jobs in heterogeneous environment.

%FIFO is provided by Hadoop and schedules jobs/tasks to the first available machines based on their releasing times. Identical-machine assumes all machines have equal processing speed, and applies Algorithm~\ref{alg1} to schedule jobs and tasks. Map-only considers only the scheduling of the map phase (using Algorithm~\ref{alg1}) while schedules reduce tasks based on FIFO. LATE is a popular scheduler for speculatively scheduling jobs in heterogeneous environment. Finally, MarS~\cite{yuan2014joint} considers the joint scheduling of map and reduce tasks to minimize total weighted completion time for a special case with identical machine processing speed. We will evaluate S-DMRS  through both experiments on a Hadoop cluster and a trace-driven simulation with online re-optimization using the Google trace~\cite{GoogleTrace}.

{\noindent \bf Evaluation setup:} We set up a heterogeneous cluster. The cluster contains $12$ (virtual) machines, and each machine consists of a physical core and 8GB memory. Each machine can process one task at a time. In the cluster, machines are connected to a gigabit ethernet switch and the link bandwidth is 1Gbps. The heterogeneous cluster contains two types of machines, fast machines and slow machines. The processing speed ratio between a fast machine and a slow machine is $8$. We evaluate S-PC by using three benchmarks -- WordCount, WordMean, and Sort. Both WordCount and WordMean are a CPU-bound application, and Sort is an I/O-bound application. Our input data is extracted from Facebook dataset. As shown in Figure~\ref{fig:inputData}, the difference between maximum task execution time and minimum task execution time is 8 sec, and it is $24\%$ of the mean. The number of reduce tasks per job is set based on workload of the reduce phase. We set the number of reduce tasks per job in WordCount and WordMean to be 1, and in Sort to be 4. All jobs are associated with weights, and values of weights are uniformly distributed between 1 to 5. Also, all jobs are partitioned into two releasing groups, and each group contains the same number of jobs. The releasing time interval between two groups is $60$sec. The completion time of a job is measured by the hour.

%\begin{figure*}[!t]
%	\centering

%	\begin{subfigure}[b]{0.32\textwidth}
%		\includegraphics[height=1.7in,width=\textwidth]{./figures1/figure1N.pdf}%
		% \vspace*{-.2cm}
%		\label{fig:ex1}%
%			\caption{Total weighted completion time of jobs with identical task sizes and workloads.}
%	\end{subfigure}
%\begin{subfigure}[b]{0.32\textwidth}
%		\includegraphics[height=1.7in, width=\textwidth]{./figures1/figure2N.pdf}%
		%    \vspace*{-.2cm}
%		\label{fig:ex2}%
%		\caption{Total weighted completion time with heterogeneous task sizes and workloads.}
%	\end{subfigure}
%\begin{subfigure}[b]{0.32\textwidth}
%		\includegraphics[height=1.7in,width=\textwidth]{./figures1/figure3N.pdf}%
		%    \vspace*{-.2cm}
%		\label{fig:ex3}%
%		\caption{ Total weighted completion time with increased task heterogeneous task sizes.}
%	\end{subfigure}
%	\caption{Comparing total weighted completion time under various scenarios}
%	\label{fig:x}	
%\end{figure*}

\begin{figure*}[]
\begin{minipage}{0.32\textwidth}
\begin{center}
		\includegraphics[height=1.7in,width=\textwidth]{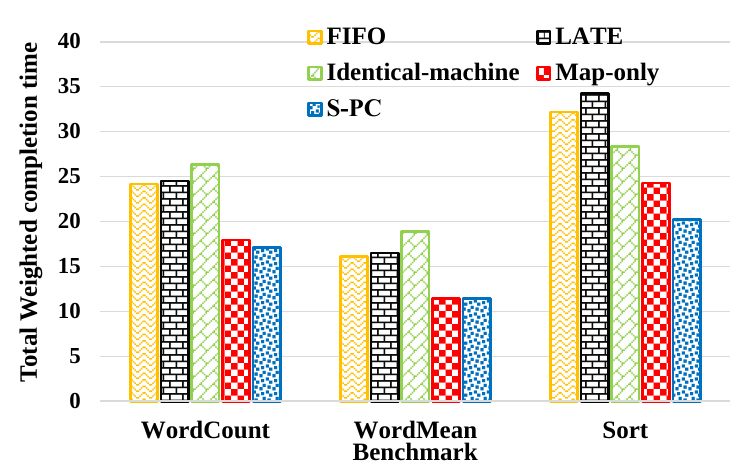}
	\caption{Total weighted completion time of jobs with identical task sizes and workloads.}
	\label{fig:ex1}
\end{center}
\end{minipage}
\hspace{0.1cm}
\begin{minipage}{0.32\textwidth}
\begin{center}
		\includegraphics[height=1.7in, width=\textwidth]{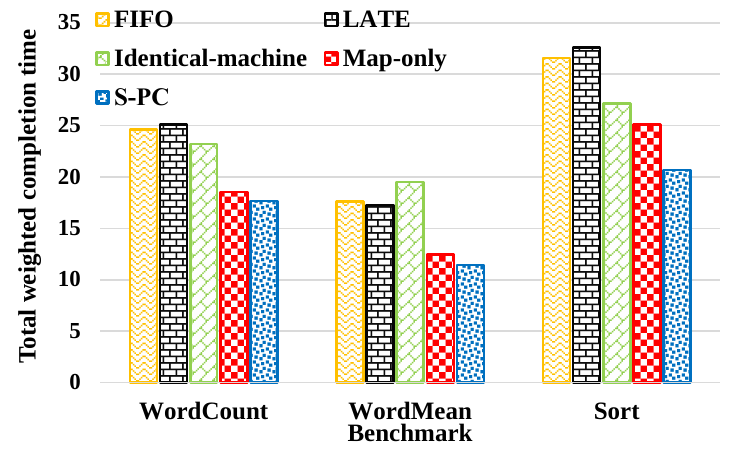}%
		\caption{Total weighted completion time with heterogeneous task sizes and workloads.}
				\label{fig:ex2}%
\end{center}
\end{minipage}
\hspace{0.1cm}
\begin{minipage}{0.32\textwidth}
\begin{center}
		\includegraphics[height=1.7in,width=\textwidth]{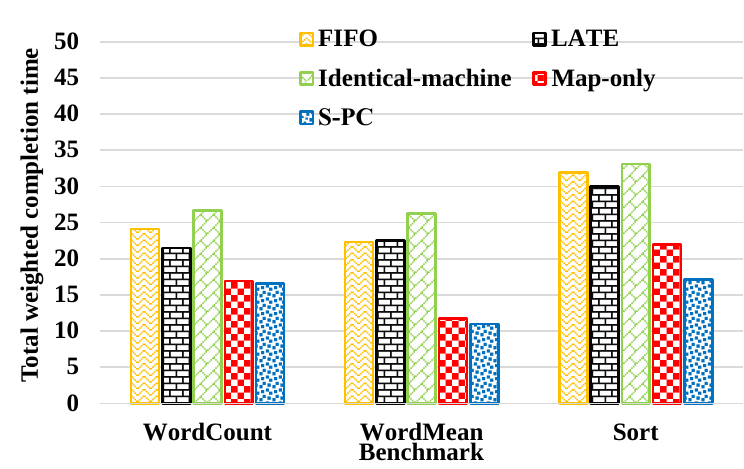}%
		   %\vspace*{-.5cm}
		\caption{ Total weighted completion time with increased task heterogeneous task sizes.}
	\label{fig:ex3}
\end{center}
\end{minipage}
\vspace{-.2in}
\end{figure*}

{\noindent \bf Experiment results:} In the first set of experiments, each experiment contains 20 jobs, and the workload of a job is 2GB. The task sizes of all jobs are the same, and equal 64MB. Figure~\ref{fig:ex1} shows that S-PC outperforms FIFO, LATE, Identical-machine, and Map-only by up to $37\%$, $38\%$, $28\%$, and $16\%$ respectively. In particular, 
%FIFO schedules jobs based on the jobs' release order. Also, computation resources can be acquired by a job until the job send a resource request to scheduler. So, jobs with high weights (time-sensitive jobs) cannot be scheduled first, so time-sensitive jobs cannot be completed in time. For task scheduling, FIFO schedules tasks to the first available container, and does not consider the heterogeneous cluster environment. Such task scheduling scheme can increase the completion time of tasks, since a container which becomes available later might be launched on a fast machine and be able to complete a task faster. 
LATE launches extra copies for tasks whose progress speed is much slower than others. Even though introducing extra copies can reduce task completion time, they also consume extra cloud resources, which otherwise can be assigned to other tasks or other jobs. 
%Also, extra copies might be scheduled after all tasks, since priority of an extra copy is lower than task's. Given a job with large number of tasks, extra copies can not speed up task's progress, but can only consume extra computation resources.
Furthermore, LATE records machines' processing speeds, and tries to avoid scheduling tasks on slow machines, so resources on slow machines are often under-utilized. LATE might schedule reduce tasks soon after the job is scheduled, and before the last map task is scheduled. Even though such scheme leaves more time for reduce tasks to fetch data from map tasks' outputs, reduce tasks can start to process data until all map tasks finish, and reduce tasks might occupy containers on fast machines, which can be assigned to other map tasks. Sort benchmark has heavy reduce workload, and reduce tasks in LATE waste more cluster resources.
%Identical-machine distributes the same number of tasks to each machine. The job's completion time is determined by the tasks' completion time running on slow machines. Also, after finishing processing their workload, fast machines need to wait for slow machines to finish their tasks. 
Map-only schedules jobs and tasks without considering the reduce phase. Under benchmarks with light workloads in the reduce phase, \eg under WordCount and WordMean benchmarks, Map-only can achieve comparable performance as S-PC. However, under benchmarks with heavy workloads in reduce phase, \eg under Sort, scheduling reduce tasks to random machines result in performance degradation and increase total weighted completion time. S-PC schedules jobs based on their weights, average task completion time, and machine speeds to minimize total weighted completion time. Also, S-PC assigns a task to a machine, which can finish the task earliest, and such task scheduling scheme can evenly distribute workloads to all machines and fully utilize cluster resources.

Figure~\ref{fig:ex2} shows the results of evaluating S-PC, in terms of total weighted completion time, by employing jobs with different task sizes and different amount of workloads. In the set of experiments, each experiment contains 20 jobs. Of these 20 jobs, 12 jobs need to process 2GB data each, and the task size is 64MB; 4 jobs need to process 1GB each, with a task size of 32MB; and the remaining 4 jobs need to process 4GB, with a task size of 128MB. Figure~\ref{fig:ex2} shows that S-PC outperforms FIFO, LATE, Identical-machine, and Map-only by up to $35\%$, $36\%$, $40\%$, and $18\%$, respectively. FIFO and LATE schedule reduce tasks soon after jobs are scheduled, and those reduce tasks occupy fast machines, which can be assign to map tasks. So, S-PC outperforms FIFO and LATE by up to $35\%$ and $36\%$, respectively. Also, FIFO and LATE do not try to minimize weighted completion time, jobs with small workloads and large weights might be scheduled after jobs with large workloads and small weights. 

We further increase the number of large jobs and small jobs, and set the number of jobs in each experiment to be 18 to make the total workload of jobs in each experiment be roughly the same as the first two sets of experiments'. Of the 18 jobs, 6 jobs need to process 2GB data each, with a task size of 64MB; another 6 jobs need to process 1GB each, with a task size of 32MB; and the remaining 6 jobs need to process 4GB each, with a task size of 128MB. Figure~\ref{fig:ex3} shows that total weighted completion time of S-PC does not grow much as the number of large jobs increases, since large jobs with small weights can be scheduled after jobs with small workloads and large weights.

%\begin{figure*}[]
%\begin{minipage}{0.32\textwidth}
%\begin{center}
%		\includegraphics[height=1.2in,width=\textwidth]{./figures1/figure4N.pdf}
%	\caption{Weighted completion time with different ratios of large/small jobs.}
%	\label{fig:y}
%\end{center}
%\end{minipage}
%\hspace{0.1cm}
%\begin{minipage}{0.32\textwidth}
%\begin{center}
%		\includegraphics[height=1.2in, width=\textwidth]{./figures1/SimulationResult1.pdf}%
%		\caption{Comparing S-DRS and MarS via Google-trace simulation and online re-optimization.}
%				\label{fig:sim1}%
%\end{center}
%\end{minipage}
%\hspace{0.1cm}
%\begin{minipage}{0.32\textwidth}
%\begin{center}
%		\includegraphics[height=1.2in, width=\textwidth]{./figures1/SimulationResult2.pdf}%
		   %\vspace*{-.5cm}
%		\caption{The evolution of accumulative weighted completion time (over time $t$) in the trace-driven simulation.}
%	\label{fig:sim2}%
%\end{center}
%\end{minipage}
%\vspace{-.2in}
%\end{figure*}

\begin{figure}[!t]
	\centering
	\includegraphics[width=0.48\textwidth]{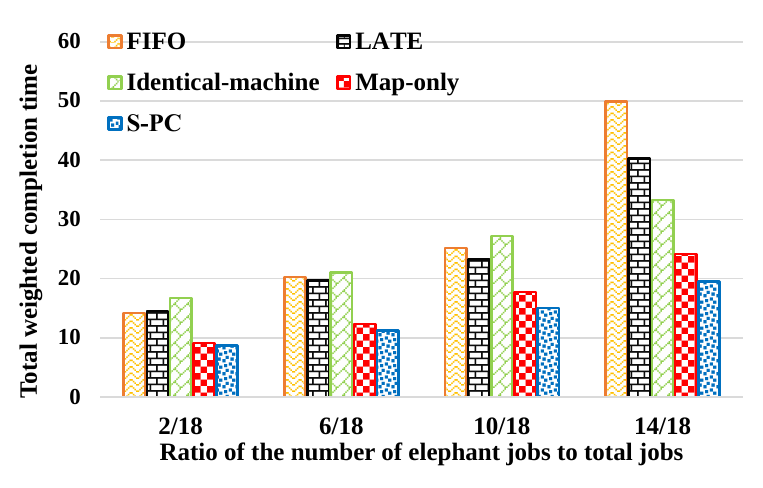}
	\caption{Weighted completion time with different ratios of large/small jobs.}
	\label{fig:y}
	\vspace{-.15in}
\end{figure}

Next, we fix the number of WordCount jobs, \ie 18 jobs, and change the number of large jobs. We set the task size to be 128MB. An large job needs to process 4GB data, and a small job needs to process 1GB data. Figure~\ref{fig:y} shows that S-PC outperforms FIFO, LATE, Identical-machine, and Map-only by up to $60\%$, $50\%$, $37\%$, and $20\%$ respectively. As the number of large jobs increases, the total weighted completion time increases. Under FIFO and LATE, as the number of large jobs increases, small jobs with large weights might be scheduled after more large jobs, and this results in long waiting times for small jobs, causing large increase in total weighted completion time. Also, as the number of large jobs increases, the increments of Identical-machine and Map-only are much larger than S-PC's. For Identical-machine, the difference of the amount of time used to complete all assigned tasks on fast machines and on slow machines increases, as the number of large jobs increases. For Map-only, as total workload in reduce phase increases, the scheduling of reduces tasks is more important for minimizing total weighted completion time. We observe that in heterogeneous environment, S-PC can significantly improve the total weighted completion time by optimally placing and scheduling all tasks with respect to their processing time and machine speeds, while simply launching extra copies for slow-running tasks (similar to LATE), or considering only map phase (similar to Map-only) are inadequate.

\section{Conclusions and Future Work} \label{sec:conclusions}

This paper considers the related machine scheduling problem for minimizing weighted sum completion time under arbitrary precedence constraints and on heterogeneous machines with different processing speeds. The precedence between any pair of tasks is modeled through a DAG, and an efficient algorithm is proposed to solve the optimization with an approximation ratio of $2(1+(m-1)/D)$ for zero release time and $2(1+(m-1)/D)+1$ for general release times. Our results significantly improve prior work - $O(\log m/\log \log m)$ in \cite{Li:2017} (which is only for zero release times) - and achieves nearly optimal performance when the number of tasks to schedule is sufficiently larger than the number of machines available. We implement the proposed scheduling algorithm and evaluate its performance in Hadoop. The numerical results show up to $82\%$ improvement over several baselines in terms of total weighted completion time.

{ We note that this paper did not account for the communication delay time when two jobs with precedence constraints are run on different machines like in \cite{davies2021scheduling}; such an extension is an important future direction. Further, the proposed analysis does not lead to effiicient bound for small $D$, and having better analytical result that achieves the bound in \cite{Li:2017} for small $D$  and our bound of $O(1)$ for large $D$ is an open problem. Online algorithms akin to that in \cite{gupta2020greed} are another future direction. }
	
%\section{Acknowledgment}

%This work is supported by ONR Grant N00014-20-1-2146 and CISCO research gift 1155109.
	
%The precedence constraint between the map tasks and the reduce tasks in MapReduce jobs is captured to give a scheduling algorithm that optimizes the weighted completion time of all jobs. The problem is NP-hard and the proposed solution uses scheduling of different tasks on the servers using a solution of a linear program, that can be solved in polynomial time. The proposed approach is shown to be approximately optimal, with a competitive ratio of $2(1+(m-1)/D)+1$, where $m$ is the number of servers and $D\ge 1$ is the task-skewness product. The competitive ratio  is shown to be $2(1+(m-1)/D)$ when all the jobs are released at time $0$. The algorithm is implemented on Hadoop framework, and  compared with other schedulers. Results demonstrate significant improvement of our proposed algorithm as compared to the baseline schedulers. 

%\vspace{-.1in}

%\newpage
%\input{7_appendix.tex}
\bibliographystyle{IEEEtran}

\bibliography{8_references,references_new,8_references1}

\begin{IEEEbiography}[{\includegraphics[width=1in,height=1.25in,clip,keepaspectratio]{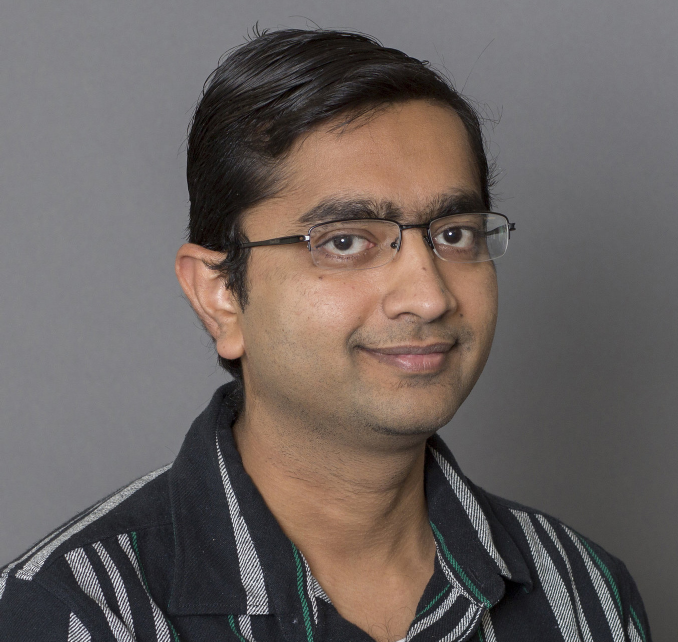}}]{Vaneet Aggarwal (S'08 - M'11 - SM'15)}
	received the B.Tech. degree in 2005 from the Indian Institute of Technology, Kanpur, India, and the M.A. and Ph.D. degrees in 2007 and 2010, respectively from Princeton University, Princeton, NJ, USA, all in Electrical Engineering.
	
He is currently an Associate Professor at Purdue University, West Lafayette, IN, where he has
been since Jan 2015. He was a Senior Member of
Technical Staff Research at AT\&T Labs-Research,
NJ (2010-2014), Adjunct Assistant Professor at
Columbia University, NY (2013-2014), and VAJRA Adjunct Professor at IISc
Bangalore (2018-2019). His current research interests are in communications
and networking, cloud computing, and machine learning.

	Dr. Aggarwal received Princeton University's Porter Ogden Jacobus Honorific Fellowship in 2009, the AT\&T Vice President Excellence Award in 2012, the AT\&T Key Contributor Award in 2013, the AT\&T Senior Vice President Excellence Award in 2014, and Purdue University's Most Impactful Faculty Innovator in 2020. He also received the 2017 Jack Neubauer Memorial Award recognizing the Best Systems Paper published in the IEEE TRANSACTIONS ON VEHICULAR TECHNOLOGY, and the 2018 Infocom Workshop HotPOST Best Paper Award. He was on the Editorial Board of the IEEE TRANSACTIONS ON GREEN COMMUNICATIONS AND NETWORKING from 2017-2020. He is currently on the Editorial Board of the IEEE TRANSACTIONS ON COMMUNICATIONS,  and the IEEE/ACM TRANSACTIONS ON NETWORKING.
\end{IEEEbiography}

\begin{IEEEbiography}[{\includegraphics[width=1in,height=1.25in,clip,keepaspectratio]{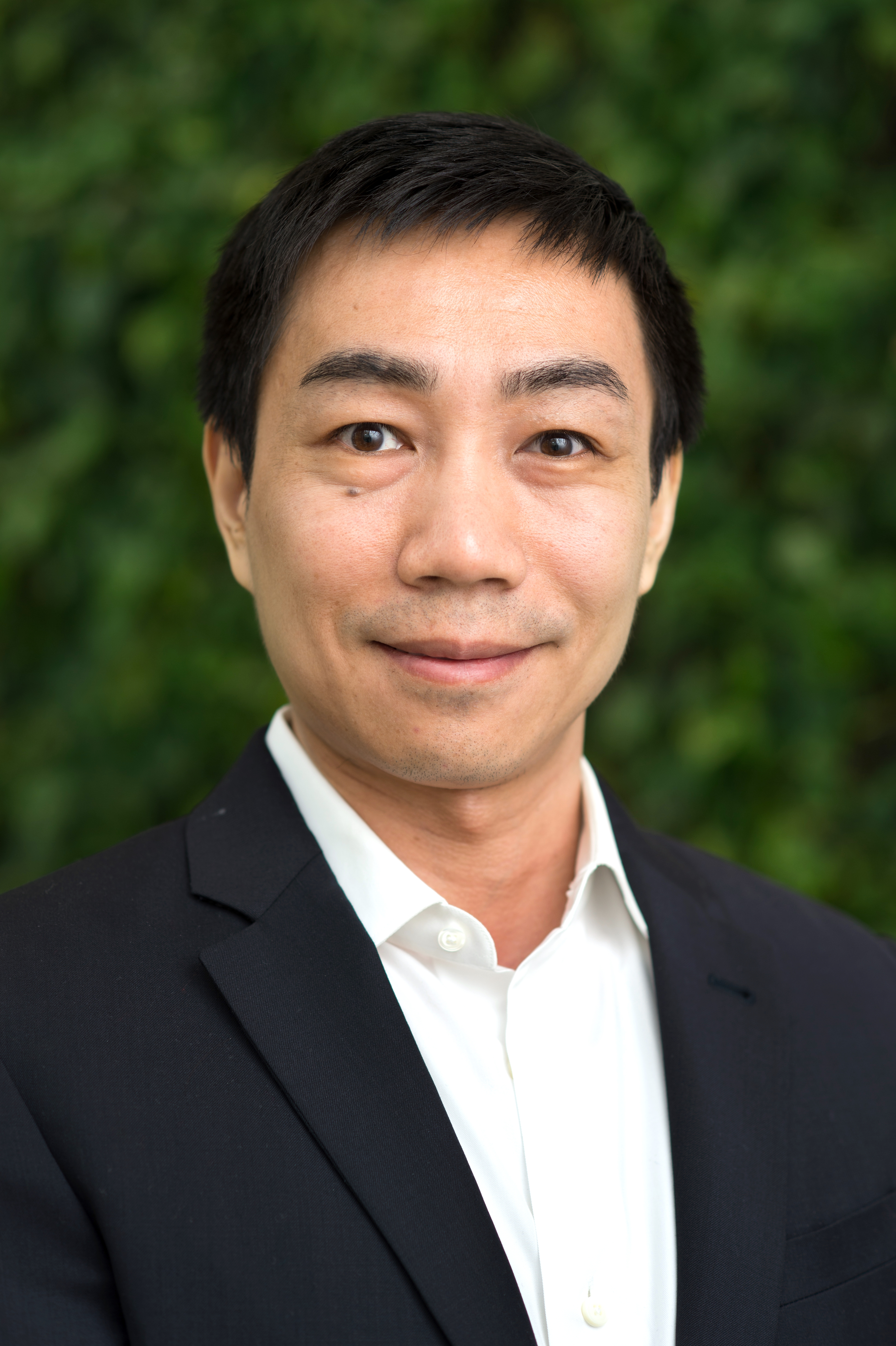}}]{Tian Lan} received the B.A.Sc. degree from the Tsinghua University, China in 2003, the M.A.Sc. degree from the University of Toronto, Canada, in 2005, and the Ph.D. degree from the Princeton University in 2010. Dr. Lan is currently an Associate Professor of Electrical and Computer Engineering at the George Washington University. His research interests include network optimization, machine learning, and network security. Dr. Lan received the SecureComm Best Paper Award in 2019, the SEAS Faculty Recognition Award at GWU in 2018, the Hegarty Faculty Innovation Award at GWU in 2017, AT\&T VURI Award in 2014, the INFOCOM Best Paper Award in 2012, the IEEE GLOBECOM Best Paper Award in 2009, and the IEEE Signal Processing Society Best Paper Award in 2008.
\end{IEEEbiography}

\begin{IEEEbiography}[{\includegraphics[width=1in,height=1.25in,clip,keepaspectratio]{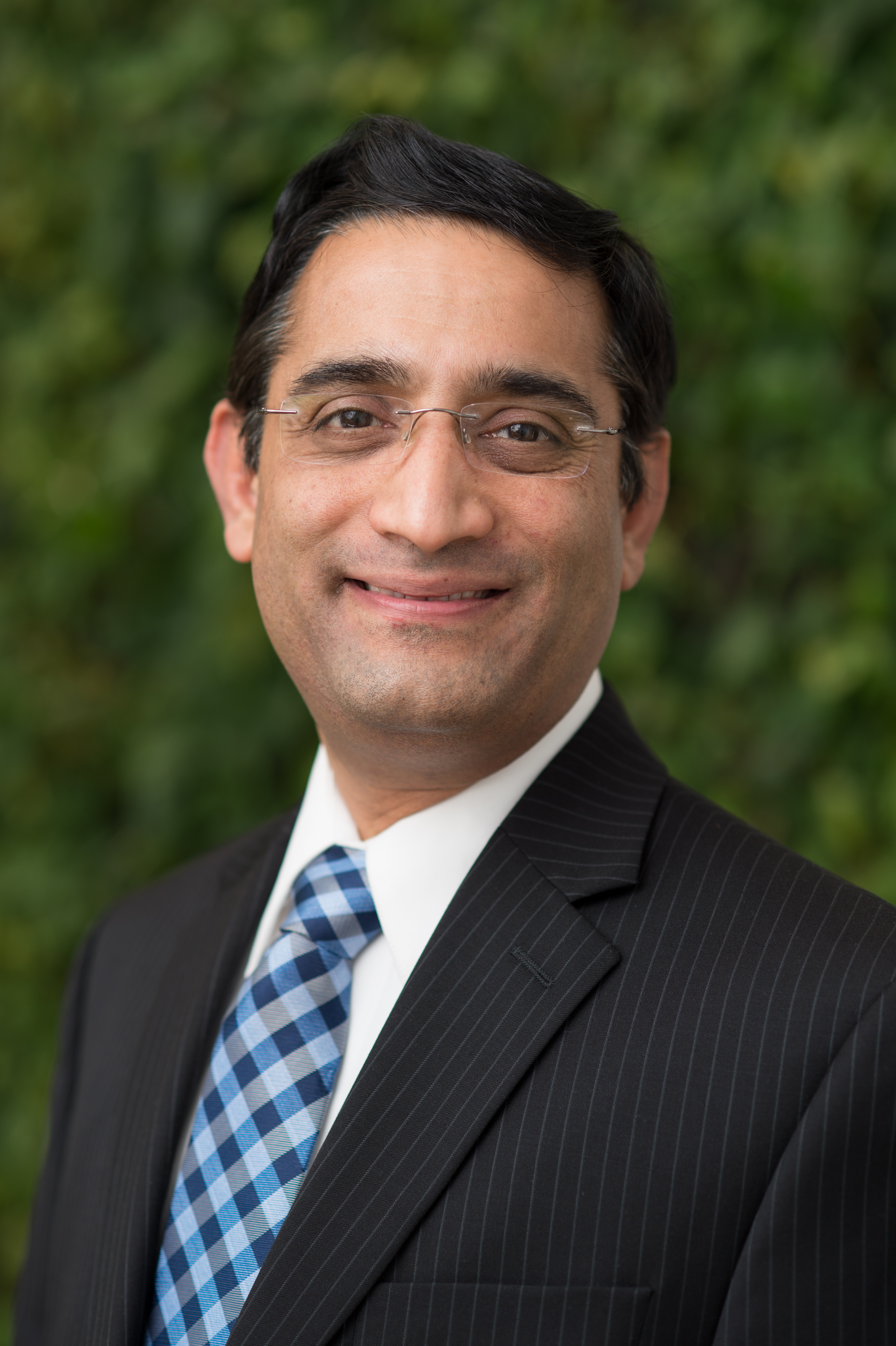}}]{Suresh Subramaniam (S'95-M'97-SM'07-F'15)}  received the Ph.D. degree in electrical engineering from the University of Washington, Seattle, in 1997. He is Professor and Chair of Electrical and Computer Engineering at the George Washington University, Washington DC, where he directs the Lab for Intelligent Networking and Computing. His research interests are in the architectural, algorithmic, and performance aspects of communication networks, with current emphasis on optical networks, cloud computing, data center networks, and IoT. He has published over 230 peer-reviewed papers in these areas.

Dr. Subramaniam is a co-editor of three books on optical networking. He has served in leadership positions for several top conferences including IEEE ComSoc’s flagship conferences of ICC, Globecom, and INFOCOM. He serves/has served on the editorial boards of 7 journals including the IEEE/ACM Transactions on Networking and the IEEE/OSA Journal of Optical Communications and Networking. During 2012 and 2013, he served as the elected Chair of the IEEE Communications Society Optical Networking Technical Committee. He has received 5 Best Paper Awards, and received the 2017 SEAS Distinguished Researcher Award from George Washington University. He is an IEEE Distinguished Lecturer during 2018-2021. He is a Fellow of the IEEE.
\end{IEEEbiography}

\begin{IEEEbiography}[{\includegraphics[width=1in,height=1.25in,clip,keepaspectratio]{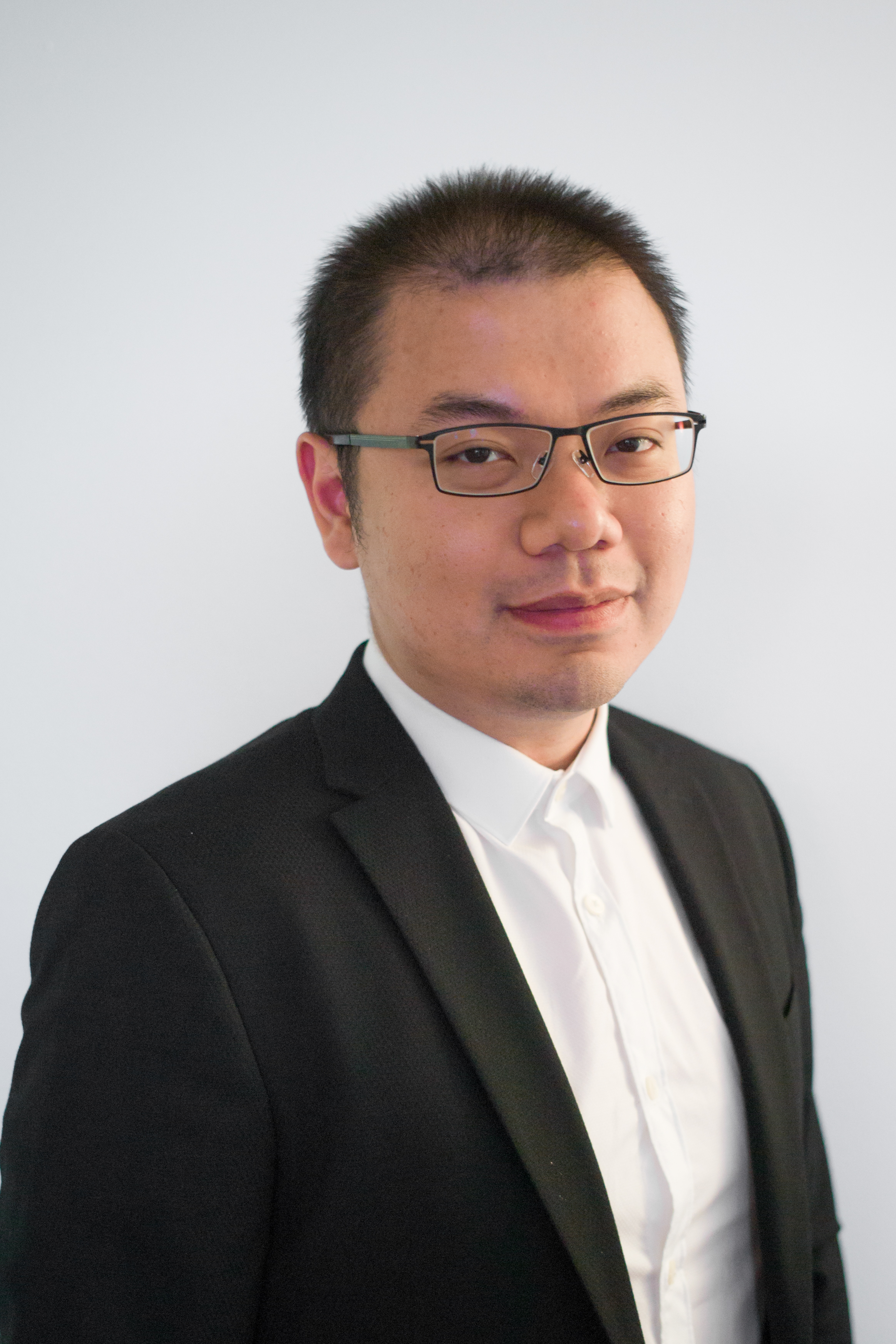}}]{ Maotong Xu}   received the B.S. degree from the Northwestern Polytechnical University, China in 2012, and the M.S. and the Ph.D. degree from the George Washington University in 2014 and in 2019, respectively. Dr. Xu is currently a senior research scientist in Facebook. His interests include real-time processing system development and optimization, and cloud computing.
\end{IEEEbiography}
\end{document}